\documentclass[12pt, draftclsnofoot, onecolumn]{IEEEtran}

\usepackage[T1]{fontenc}
\usepackage{hyperref}
\usepackage{multirow}
\usepackage{diagbox}
\usepackage{cite}
\usepackage[pdftex]{graphicx}
\usepackage{graphicx, subfigure}
\usepackage{amsmath}
\usepackage[cmintegrals]{newtxmath}

\begin{document}

\title{Convolutional Neural Network based Multiple-Rate Compressive Sensing for Massive MIMO CSI Feedback: Design, Simulation, and Analysis}

\author{\normalsize {Jiajia~Guo, Chao-Kai~Wen, \IEEEmembership{\normalsize {Member,~IEEE}},~Shi~Jin, \IEEEmembership{\normalsize {Senior Member,~IEEE}},\\~and Geoffrey~Ye~Li, \IEEEmembership{\normalsize {Fellow,~IEEE}}}

\thanks{J.~Guo and S.~Jin are with the National Mobile Communications Research
Laboratory, Southeast University, Nanjing, 210096, P. R. China (email: jjguo@mail.ustc.edu.cn, jinshi@seu.edu.cn).}
\thanks{C.-K.~Wen is with the Institute of Communications Engineering, National Sun Yat-sen University, Kaohsiung 80424, Taiwan (e-mail: chaokai.wen@mail.nsysu.edu.tw).}
\thanks{G. Y. Li is with the School of Electrical and Computer Engineering, Georgia Institute of Technology, Atlanta, GA 30332 USA (e-mail: liye@ece.gatech.edu).}
}

\maketitle

\begin{abstract}
Massive multiple-input multiple-output (MIMO) is a promising technology to increase link capacity and energy efficiency.
However, these benefits are based on available channel state information (CSI) at the base station (BS).
Therefore, user equipment (UE) needs to keep on feeding CSI back to the BS, thereby consuming precious bandwidth resource.
Large-scale antennas at the BS for massive MIMO seriously increase this overhead.
In this paper, we propose a multiple-rate compressive sensing neural network framework to compress and quantize the CSI. This framework not only improves reconstruction accuracy but also decreases storage space at the UE, thus enhancing the system feasibility.
Specifically, we establish two network design principles for CSI feedback, propose a new network architecture, CsiNet+, according to these principles, and develop a novel quantization framework and training strategy.
Next, we further introduce two different variable-rate approaches, namely, SM-CsiNet+ and PM-CsiNet+, which decrease the parameter number at the UE by 38.0\% and 46.7\%, respectively.
Experimental results show that CsiNet+ outperforms the state-of-the-art network by a margin but only slightly increases the parameter number.
We also investigate the compression and reconstruction mechanism behind deep learning-based CSI feedback methods via parameter visualization, which provides a guideline for subsequent research.

\end{abstract}


%
\IEEEpeerreviewmaketitle

\section{Introduction}
\label{introduction}

\IEEEPARstart{M}{assive} multiple-input multiple-output (MIMO) is a critical technology for 5G and beyond systems\cite{6736746,marzetta2015massive,wong2017key}.
In massive MIMO systems, base stations (BSs), equipped with a large number of antennas, can recover information received from user equipment (UE) at low signal-to-noise-ratio (SNR) and simultaneously serve multiple users\cite{larsson2014massive,6798744,6457363}. 
However, BSs should obtain the instantaneous channel state information (CSI) to acquire these potential benefits and the accuracy of the obtained CSI directly affects the performance of the massive MIMO systems\cite{larsson2014massive}.
For the uplink, BSs can easily estimate CSI accurately through the pilots sent by the UE.
However, the downlink CSI is difficult to achieve, especially in frequency-division duplexing (FDD) systems, which are employed by the most cellular systems nowadays.
In time-division duplexing (TDD) systems, downlink CSI can be inferred from uplink CSI utilizing the reciprocity\cite{7727995}.
However, in FDD systems, weak reciprocity is present , thereby making it hard to infer downlink CSI by observing uplink CSI\cite{sim2016compressed}.

In traditional MIMO systems, downlink CSI in FDD systems is first estimated at the UE by the pilots and then fed back to the BS.
However, this feedback strategy is infeasible in massive MIMO because the substantial antennas at the BS greatly increase the dimension of CSI matrix, thereby leading to a large overhead\cite{4641946,7727995}.
To address this issue, the CSI matrix should be efficiently compressed\cite{4641946,qin2018sparse}, which can be based on compressive sensing (CS) or deep learning (DL).
The CS-based methods exploit the sparsity of massive MIMO CSI in certain domain\cite{8284057}.
In \cite{6214417}, CS has been first applied to CSI feedback in the spatial-frequency domain, which exploits the high spatial correlation of CSI resulting from the limited distance among antennas in massive MIMO. 
In \cite{6816089}, a hidden joint sparsity structure in the user channel matrices has been found and exploited due to the shared local scatterers.
CS techniques simplify the encoding (compression) process; but, the decoding (decompression) process turns into solving an optimization problem and  demands substantial computing sources and time\cite{qin2018sparse}, thereby making it difficult to implement in many practical communication systems.

DL recently made tremendous strides in several aspects, including computer vision and natural language processing\cite{lecun2015deep,young2018recent}. 
The DL-based non-iterative methods have shown outstanding performance on image compression.
Traditional algorithms use iterative methods to solve image reconstruction optimization problem.
The ReconNet in \cite{kulkarni2016reconnet} recovers images utilizing stacked convolutional layers without iteration, thereby reducing the reconstruction time by a margin. 
From \cite{yao2017dr2}, it is a good strategy to generate a preliminary reconstruction at the decoder via a linear mapping network and then use a residual network to refine estimates.
In \cite{cui2018deep}, the random Gaussian measurement matrix is replaced by a learned measurement matrix at the encoder. 

The DL-based image compression technique is first introduced to massive MIMO CSI feedback in \cite{8322184} based on the autoencoder architecture in \cite{vincent2008extracting}. In the CsiNet in \cite{8322184}, the encoder acts as the role of compression module instead of randomized Gaussian matrix at the UE and the decoder is regarded as the decompression module at the BS.
Then, CsiNet-LSTM in \cite{8482358} improves the reconstruction accuracy by considering the temporal correlations of CSI utilizing long-short time memory (LSTM) architecture\cite{greff2017lstm}.
From \cite{8543184}, this neural network architecture can be modified to significantly reduce the number of network parameters.
Based on the reciprocity between bi-directional channels, uplink CSI information can help reconstruct the downlink CSI in \cite{8638509}.
The work in \cite{jang2019deep} reduces impact of the feedback transmission errors and delays.
It has been shown in \cite{wu2019compressed} that the performance of DL-based measurement matrix is better than that of randomized measurement matrices (e.g., Gaussian and Bernoulli distribution).

In the above-mentioned DL-based work, DL-based models are regarded as a black box and have no interpretation of why excellent performance can be obtained.
Meanwhile, the impact of the quantization process has been ignored, thereby leading to substantial errors in practical wireless communication systems.
The CSI feedback in massive MIMO systems should be drastically compressed while the coherence time is short and vice versa.
Therefore, the compression rate ($CR$) must be adjusted according to the environments.
The iterative algorithms should be able to work for different $CR$s.
However, the existing DL-based methods can only compress the CSI matrix with a fixed $CR$.
The UE has to store several CS network architectures and corresponding parameter sets to realize multiple-rate CSI compression, which is infeasible due to the limited storage space at the UE.

\begin{figure*}[t]
    \centering 
     \includegraphics[scale=0.58]{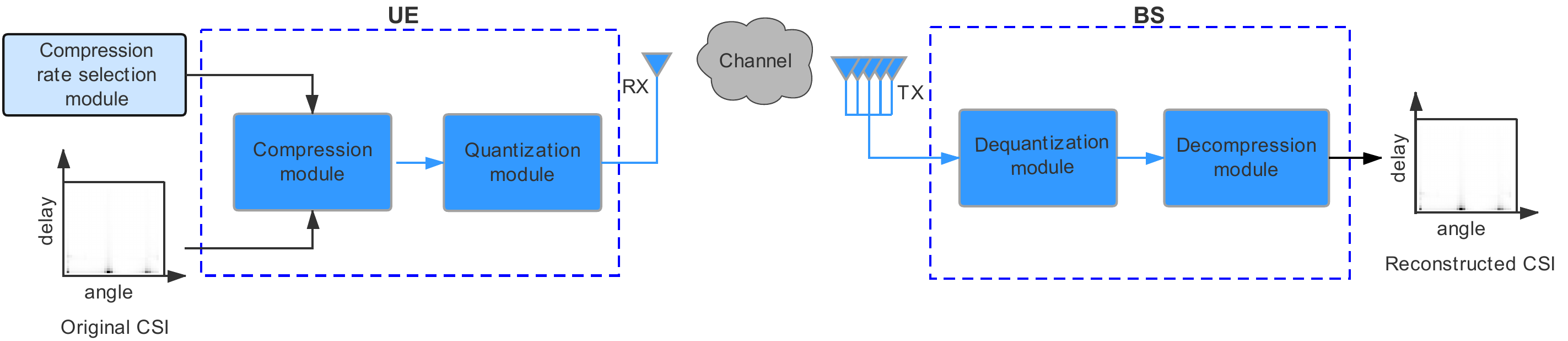}    
	\caption{\label{overview1}Overview of the multiple-rate bit-level compressive sensing CSI feedback framework. UE compresses CSI matrix with a selected $CR$, quantizes the measurement vector, and then transmits it. Once BS receives the transmitted bitstream, it dequantizes bitstream and then decompresses measurement vector. } 
\end{figure*}

In this work, we propose a multiple-rate compressive sensing framework as shown in Fig. \ref{overview1}, which will not only improves the reconstruction accuracy but also bridges the gap between DL-based methods and practical deployment.  
First, we introduce a new network architecture, namely, CsiNet+, modified from CsiNet, which exploits the sparsity characteristics of CSI in angular-delay domain and the refinement theory.
Then, we develop a novel framework and training strategy for quantization, which does not need extra storage space for different quantization rates at the UE.
Subsequently, two different network frameworks are developed for variable $CR$ compression, thereby greatly saving storage space. 
Finally, we discuss the compression and reconstruction mechanism of DL-based methods via parameter visualization and evaluate the performance.

The major contributions of this work are summarized as follows:
\begin{itemize}
\item After investigating the characteristics of CSI from the aspect of sparsity and the key idea of refinement theory in DL, we propose two network design principles for CSI feedback, which provides a guideline for future network design.
We propose a new network architecture, named CsiNet+, which improves the original CsiNet.
\item We introduce a novel quantization framework and training strategy, which is especially suitable to CSI feedback in massive MIMO systems.
This framework and  training strategy require no architectural change or parameter update at the UE.
Neural networks are used to offset the quantization distortion.
Furthermore, different quantization rates can be realized without increasing parameter number or computational resource at the UE.
\item We propose two different variable rate frameworks, namely, SM-CsiNet+ and PM-CsiNet+.
They also reduce the parameter number by 38.0\% and 46.7\%, respectively, thereby greatly saving the storage space at the UE.
This work is the first to address variable $CR$ issue in DL-based CSI feedback.
\item We investigate the compression and reconstruction mechanism of DL-based CSI feedback via parameter visualization and obtain insightful understanding of DL-based CSI feedback, which is the first to reveal the reason behind the excellent performance of DL-based methods, and provides important guidelines for subsequent research in this area.
\end{itemize}

The rest of this work is organized as follows.
In Section \ref{SystemModel}, we introduce the system model, including channel model and CSI feedback process.
Then, the novel network architecture CsiNet+ and quantization framework are presented in Section \ref{BitCsiNet+}. 
Section \ref{Multiple-rate} introduces two different variable rate frameworks.
In Section \ref{results}, we provide the experiment details and numerical results of the proposed networks and frameworks, and reveal the compression and reconstruction mechanism. 
Section \ref{Conclusion} finally concludes our work.

\section{System Model}
\label{SystemModel}
After introducing the massive MIMO-OFDM system in this section, we will describe the CSI feedback process.
\subsection{Massive MIMO-OFDM system}
\label{channelsparisity}
We consider a single-cell FDD massive MIMO-OFDM system, where there are $N_t(\gg1)$ transmit antennas at the BS and a single receiver antenna at the UE, OFDM is with $N_c$ subcarriers.
The received signal at the $n$-th subcarrier can be expressed as follows:
\begin{equation}\label{csiH}
y_n = \tilde{\mathbf{h}}_n^{H} \mathbf{v}_n x_n+z_n,
\end{equation}
where $\mathbf{\tilde{h}}_n$ and $\mathbf{v}_n$ $\in\mathbb{C}^{N_t\times1} $ are the channel frequency response vector and the precoding vector at the $n$-th subcarrier, respectively, $x_n$ represents the transmitted data symbol, $z_n$ is the additive noise or interference, and $(\cdot)^ H $ represents conjugate transpose.
The CSI matrix in the spatial-frequency domain can be expressed in matrix form as $\mathbf{\tilde{H}} = [\mathbf{\tilde{h}}_1,\mathbf{\tilde{h}}_2,...,\mathbf{\tilde{h}}_{N_c}]^H\in\mathbb{C}^{N_t\times N_c}$.

In the FDD system, UE estimates the downlink channel and then feeds this information (CSI) to the BS. 
With the downlink CSI, the BS calculates precoding vector $v_n\in\mathbb{C}^{N_t\times1} $ via singular value decomposition.
The number of feedback parameters is $2N_c  N_t$, which is proportional to the number of antennas. 
Excessive feedback in massive MIMO system greatly occupies the precious bandwidth.

We consider reducing feedback overhead by exploiting the sparsity of CSI in the angular-delay domain.
The CSI matrix in the spatial-frequency domain can be converted into the angular-delay domain by 2D discrete Fourier transform (DFT) as follows:
\begin{equation}
\mathbf{H} = \mathbf{F}_{\rm d} \mathbf{\tilde{H}} \mathbf{F}_{\rm a},
\end{equation}
where $\mathbf{F}_d$ is a $N_c \times N_c$  DFT matrix and $\mathbf{F}_a$ is a $N_t \times N_t$ matrix.
Due to the sparsity of massive MIMO channel in the angular-delay domain, most elements in the delay domain are near zero and only the first $N_{c}'$ ($<N_{c}$) rows exhibit distinct non-zero values because the time delay among multiple paths only lies in a particularly limited period.
Therefore, we directly truncate the channel matrix rows to the first $N_{c}'$ rows that are with distinct non-zero values. 
Meanwhile, the channel matrix is also sparse in a defined angle domain by performing DFT on spatial domain channel vectors if the number of the transmit antennas $N_t \to \infty $\cite{6940305}.

In this paper, we regard the 2D channel matrix as an image and the normalized absolute values of CSI matrix are regarded as the gray-scale values to visualize the sparsity of the retained $N_c' \times N_t$ channel matrix $\mathbf{H}$ in the angular-delay domain, which has been demonstrated in the literature, such as \cite{liu2012cost,8354789}.

\subsection{CSI Feedback Process}
Once the channel matrix $\mathbf{H}$ in the angular-delay domain is estimated at the UE,  compression, quantization, and entropy encoding\footnote{Since entropy encoding is lossless, we do not take it into consideration in the following parts.} will be used in turn to reduce CSI feedback overhead.
The compressed CSI matrix can be expressed as follows:
\begin{equation}
\label{encoderF}
\mathbf{ H}_c =  \mathcal{Q}(f_{\rm{com}}(\mathbf{H},\Theta_1)),
\end{equation}
where $f_{\rm{com}}(\cdot)$ and $\mathcal{Q}(\cdot)$ denote the  compression and quantization processes, respectively, and $\Theta_1$ represents parameters of the compression module (encoder).

Once the BS receives the compressed CSI matrix, dequantization and decompression will be used to recover the channel matrix in the angular-delay domain,
\begin{equation}
\label{decoderF}
\mathbf{\hat H} = f_{\rm{decom}}(\mathcal{D}(\mathbf{H}_c,\Theta_2)),
\end{equation}
where $\mathcal{D}(\cdot)$ and $f_{\rm{com}}(\cdot)$ represent the dequantization and decompression functions, respectively,
and $\Theta_2$ denotes the parameters in the decompression module (decoder).
Therefore, the optimization compression and recovery can be formulated by combining (\ref{encoderF}) and (\ref{decoderF}) together with the mean-squared error (MSE) distortion metric as the following:  
\begin{equation}
\label{OptimizationGoal0}
(\hat \Theta_1, \hat \Theta_2) = \mathop{\arg\min}_{\Theta_1, \Theta_2} \ \ \|  \mathbf{H} -   f_{\rm decom}(\mathcal{D}(\mathcal{Q}(f_{\rm com}(\mathbf{H},\Theta_1))),\Theta_2)     \|_2^2.
\end{equation}

\section{CSI Compression based on Convolutional Neural Networks}
\label{BitCsiNet+}
In this section, we describe the proposed framework, which mainly includes neural network architecture, and quantization and dequantization sub-modules.

\subsection{Network Architecture for Channel Dimension Reduction}
The CsiNet in \cite{8322184}, an encoder–decoder structure, has demonstrated promising performance in CSI compression and reconstruction.
The encoder first extracts CSI features via a convolutional layer with two $3\times3$ filters, followed by an activation layer.
Then, a fully connected (FC) layer with $M$ neurons is adopted to compress the CSI features to a lower dimension.
The $CR$ of this encoder can be calculated by:
\begin{equation}
CR = \frac{2N_tN_c}{M}.
\end{equation}
At the decoder, the first layer is also a FC layer with $2N_tN_c$ neurons, and the output vector is reshaped with the same shape of the original channel matrix.
The above layers produce the initial estimate of channel matrix $\mathbf{H}$.
Then, the output is fed into two RefineNet blocks \cite{8322184}, which are designed to continuously refine the reconstruction and contain three convolutional layers and identity shortcut connections \cite{He_2016_CVPR}.
The last layer of CsiNet is a convolutional one with batch normalization (BN) \cite{ioffe2015batch} and Sigmoid activation layer, scaling the output to the $[0,1]$ range.

\begin{figure*}[t]
    \centering 
     \includegraphics[scale=0.6]{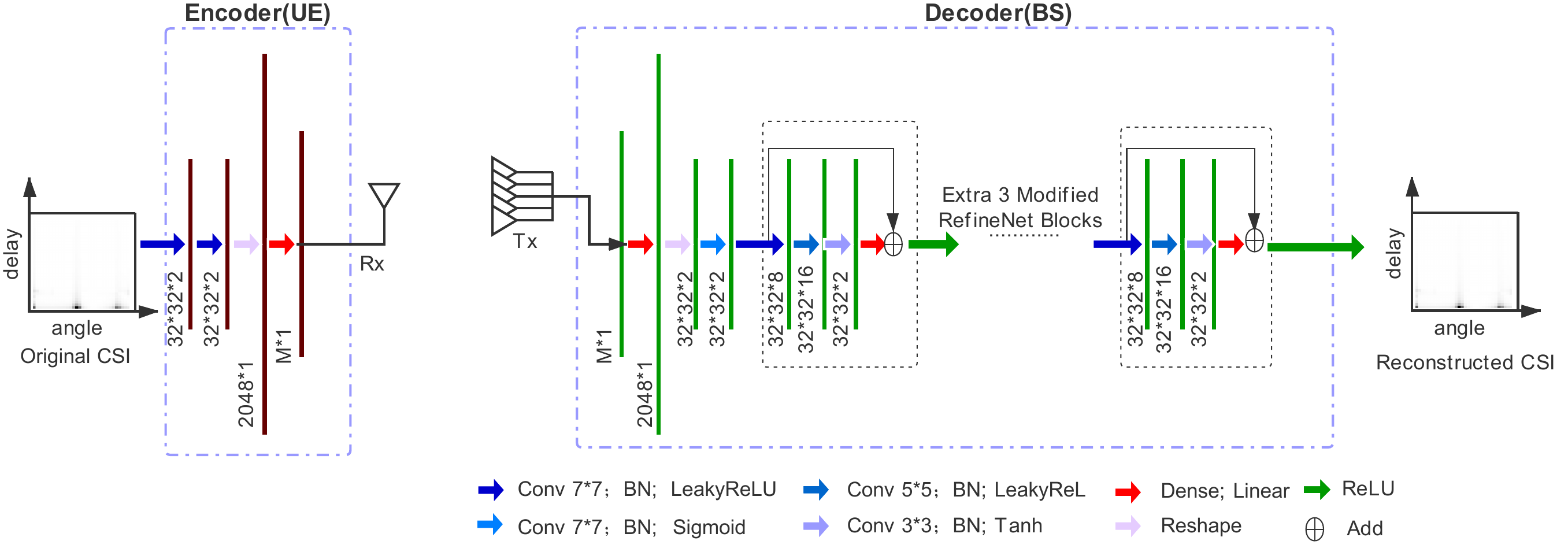}    
	\caption{\label{CsiNet+_overview}Overview of CsiNet+ architecture. The left module is an encoder at the UE, compressing the CSI matrix. Meanwhile, the right module is a decoder at the BS, reconstructing CSI matrix from the received compressive measurements. } 
\end{figure*}

The proposed architecture of neural network, CsiNet+, as shown in Fig. \ref{CsiNet+_overview}, is based on the CsiNet with two main modifications: convolutional kernel size and refinement process.
\subsubsection{Modification 1}
\label{M1}
Wireless communication channels often exhibit a block-sparse structure, that is, a matrix that exhibits nonzero values occurring in clusters\cite{eldar2009block}. 
CsiNet and other CS-based feedback methods \cite{8482358,8543184} regard this sparsity of the channel matrix as a precondition.
However, CsiNet first uses a convolutional layer with $3\times3$ filters to extract the features of block-sparse channel matrix, which is inappropriate in this scenario.
In the image scenario, $3\times3$ filters can extract the edge information within a particularly small receptive field.
In contrast, the visualization results of output of the first convolutional layer in CsiNet indicates that most output coefficients are near-zero, which is similar to the input channel matrix and contains less information.
Specifically, if the receptive field is located in a large `blank' area, the nine coefficients are still zero after being convoluted with $3\times3$ filters.
Thus, these convolution operations can be regarded as futile.
Meanwhile, the sparsity is unable to exhibit in the fully `non-blank' area.

In \cite{bai2018deep}, a new block activation unit with the block size of six has been proposed to handle the block-sparse vector in wideband wireless communication systems.
Inspired by it, we use two $7\times7$ convolutional layers in CsiNet+ to replace the first $3\times3$ convolution layer of CsiNet at the encoder.
The two series convolutional layers with $7\times7$ filters present a $13\times13$ receptive field, which is hardly located in the `blank' area due to the large convolutional kernel size.
Meanwhile, hardly no fully `non-blank' area exists, so sparsity can effectively exhibit.
For the same reason, we also replace the two series $3\times3$ convolutional layers in RefineNet block with $7\times7$ and $5\times5$ ones, which is called as CsiNet-M1.

\subsubsection{Modification 2}
\label{M2}
The key idea of the refinement is to improve the estimates from the initial ones via stacking convolutional layers and identity shortcut connections\cite{kulkarni2016reconnet}.
Each RefineNet block is optimized to ensure that its output is the same as the residual between its input and ground truth as much as possible, which can be expressed as,
\begin{equation}
\mathbf{\hat H}_{\rm{res}} = \mathbf{H} - \mathbf{\hat H}_{\rm{in}},
\end{equation}
where $\mathbf{\hat H}_{\rm{in}}$ is the initial estimate and $ \mathbf{\hat H}_{\rm{res}}$ is the expected residual.
Fundamentally, the output of the last RefineNet block should be the final estimate; otherwise, the refinement will be disturbed.
In \cite{yao2017dr2,8486521}, the RefineNet block is directly used as the last layer.
However, in CsiNet, a convolutional layer follows the last RefineNet block, thereby disturbing the refinement.
Therefore, we remove this convolutional layer in proposed CsiNet+.

Training the entire neural network DR$^2$-Net in \cite{yao2017dr2} is conducted in two steps.
First, the encoder and the first FC layer at the decoder are trained using a large learning rate to obtain a preliminary reconstructed image. 
Second, the encoder and decoder are trained jointly in an end-to-end manner using a smaller learning rate.
Obviously, the two-step training strategy is inefficient and time-consuming.
In contrast, CsiNet in \cite{8482358} is trained via end-to-end learning, where a good initial estimate $\mathbf{\hat H}_{\rm{in}}$ is difficult to obtain.
Therefore, we add a $7\times7$ convolutional layer with a BN layer and Sigmoid activation between the FC layer and the first RefineNet block considering the disadvantages of the DR$^2$-Net and the CsiNet.
The output of this added convolutional layer is regarded as the initial estimate $\mathbf{\hat H}_{\rm{in}}$, whose quality has been improved by the extra layer.
Similar to the CsiNet, the parameters of the encoder and decoder are updated during training via an end-to-end approach, which is called as CsiNet-M2.

The above are the two modifications in CsiNet+, as shown in Fig. \ref{CsiNet+_overview}.
The left module is the encoder at the UE, which compresses the CSI matrix, and the right module is a decoder at the BS, which reconstructs the CSI matrix from compressed channel. 
The loss function of CsiNet+ is MSE.

\subsection{Quantization and Dequantization}
\label{Qstrategy}
\begin{figure*}[t]
    \centering 
     \includegraphics[scale=0.62]{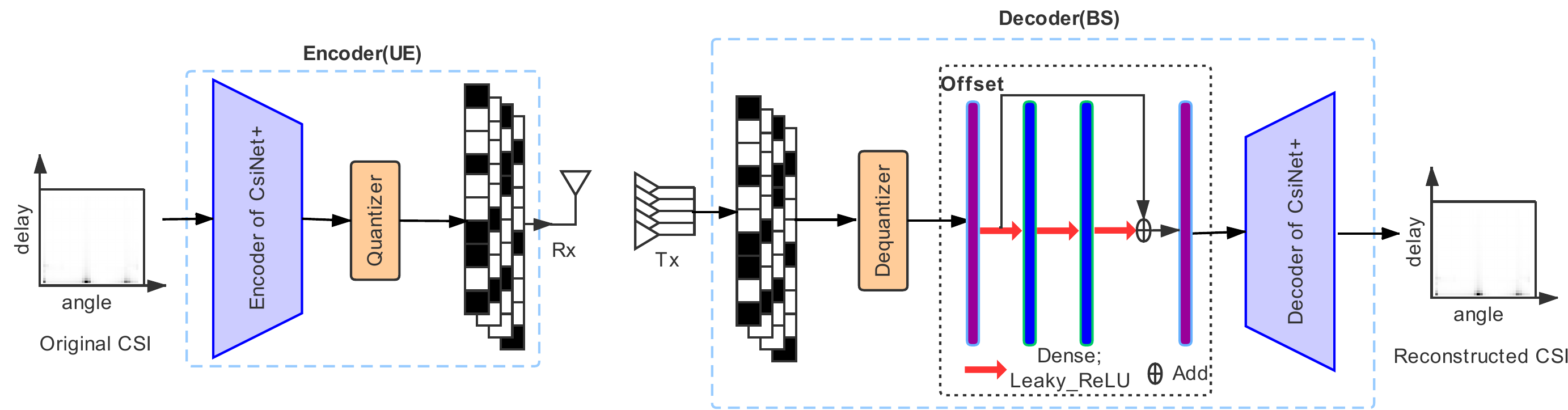}    
	\caption{\label{quantization}Proposed bit-level CsiNet+ framework. The original CSI is first compressed at the encoder (UE), and then quantization is adopted to generate a bitstream. At the decoder (BS), the received measurement vectors are first dequantized and then fed into several neural networks.} 
\end{figure*}

The output of the encoder at the UE in CsiNet+ needs to be converted into bitstream for transmission (feedback).
Therefore, the output of the CsiNet+ encoder should be first quantized.
Once the BS receives bitstream, dequantization is first used before feeding into the neural networks, as in Fig. \ref{quantization}.

\subsubsection{Quantization method}
In \cite{jang2019deep}, uniform quantization is used to discretize measurement vectors.
However, it is not optimal for compressed CSI even if uniform quantization provides good quantization for strong signal.
Therefore, in this work, we adopt  a $\mu$-law non-uniform quantizer, which is optimized by adapting a companding function $f(\cdot)$ as,
\begin{equation}
f(x) = \pm \frac{{\rm ln}(1+\mu|x|)}{1+\mu} ,
\end{equation}
where $x\in [-1,1]$ is the weak signal and $\mu$ is a constant that determines companding amplitude.

\subsubsection{Offset module}
After dequantizing the bitstream at the BS, an offset module\cite{cui2018efficient} is first used to minimize the quantization distortion as follows,
\begin{equation}
\hat {f}_{\rm off} = \mathop{\arg\min}_{f_{\rm off}}    (f_{\rm{com}}(\mathbf{H},\Theta_1)-f_{\rm off}(\mathbf{ H}_c)),
\end{equation}
where $f_{\rm off}$ denotes the offset process.
In order to model the function $f_{\rm off}$, an offset neural network is designed to minimize the distortion, as shown in Fig.\ref{quantization}. 
The offset network is based on residual learning and consisted of three FC layers in which there are $N$ neurons.

\subsubsection{Training strategy}
Since the quantization function is non-differentiable, the gradient of the entire bit-level network cannot be passed while using backpropagation learning algorithm, thereby making it impossible to train networks in the end-to-end way.
The widely used solution is to set the quantization gradient (i.e., $round(\cdot)$ gradient) to a constant.
Then, the entire neural networks including encoder and decoder are trained in an end-to-end way.
This training strategy solves the gradient backpropagation problem due to the quantization function,  but the network can only work with specific quantization bits.
If requiring different quantization bit rates, different neural networks are required and substantial parameters need storing, which is inapplicable in CSI feedback due to the limited storage space in the UE.
Occupying great storage space just for different quantization bit rates is not worthy (the number of encoder parameters sometimes is over 1M).

In contrast with the existing quantization training strategy \cite{cui2018efficient,sun2016deep,cai2018deep} that jointly trains the encoder and decoder with a fix quantization bit rate, we do not always train networks via end-to-end learning or train different encoders for different quantization bit rates.
Specifically, we first train CsiNet+ without quantization via an end-to-end approach with a large learning rate.
Next, we use non-uniform quantization to discrete the measurement vectors and the dequantizer at the BS conducts the inverse operation of quantization to recover the vectors, which generates the training set of the offset network.
Then, the offset network is optimized by the Adam optimizer\cite{kingma2014adam} with MSE loss function. 
Once the offset network is trained, we fix bit-level CsiNet+'s parameters of the encoder and fine-tune the offset and decoder networks with a small learning rate to further minimize the quantization distortion effect, which can be formulated as follows,
\begin{equation}
\label{OptimizationGoal}
(\hat \Theta_2,\hat \Theta_3 )= \mathop{\arg\min}_{\Theta_2,\Theta_3} \ \ \|  \mathbf{H} -   f_{\rm off}(f_{\rm{decom}}(\mathcal{D}(\mathcal{Q}(f_{\rm{com}}(\mathbf{H},\hat \Theta_1)),\Theta_3)),\Theta_2)\|_2 ^2,
\end{equation}
where $\hat \Theta_1$ and $\Theta_3$ denote the learned parameters at the encoder in CsiNet+ and the parameters in the offset network, respectively.
Therefore, the UE only needs to store one parameter set regardless of quantization bit rates.
The problem of different quantization bit rates is solved at the BS by training distinct decoders for different quantization bit rates, which is feasible due to the great storage space at the BS.

\section{Multiple-rate CSI Feedback}
\label{Multiple-rate}
\begin{figure*}[!t]
    \centering 
     \includegraphics[scale=0.52]{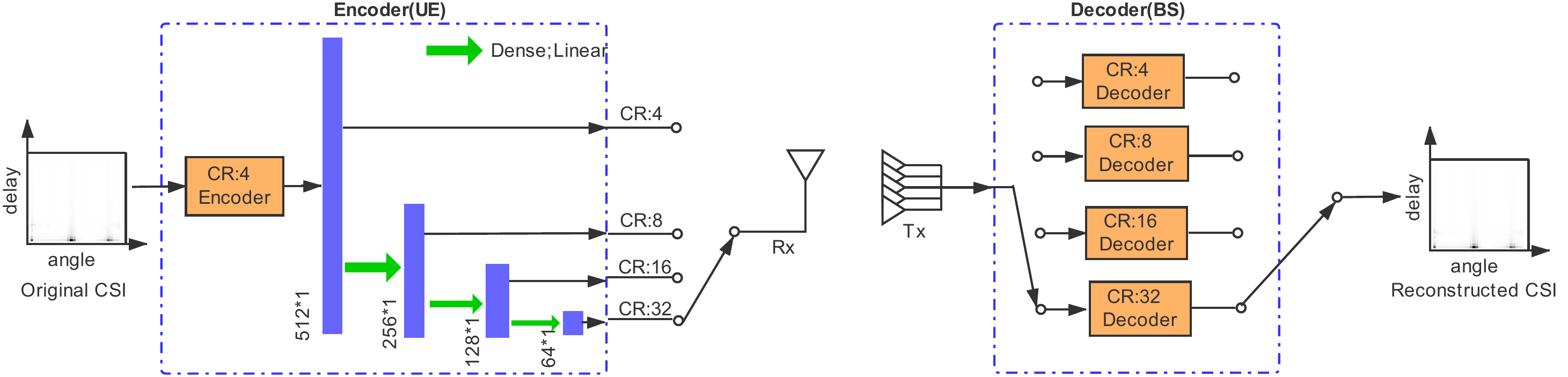}    
	\caption{\label{series}Series multiple-rate compression framework. The key idea of SM-CsiNet+ is that high compression measurement vectors can be generated from the low ones.} 
\end{figure*}
Although CSI compression can reduce feedback overhead, accuracy of reconstructed CSI at the BS is sacrificed, which may adversely affect MIMO communication network performance.
Hence, communication systems sometime need adjust the $CR$ according to the environments, as mentioned in Section \ref{introduction}.
In contrast with the traditional iterative algorithms that can work with different $CR$s, the existing DL-based methods can only compress CSI matrix with a fixed $CR$ and have to train and store a different neural network for a different $CR$, thereby occupying large storage space at the UE.
In this part, we focus on a multiple-rate framework, which can compress the CSI matrix at different $CR$s to save the storage space at the UE.
As before, we neglect the decoder parameter number at the BS because the storage space of the BS is enough.
\begin{table}[b]
\caption{Parameter numbers of CsiNet+ encoders with different $CR$s.}
\label{NumMR}
\centering
\begin{tabular}{c|c c c c}
\hline \hline
\diagbox{Number}{$CR$} & 4&8&16&32\\
\hline
Total &1,049,500&524,956&262,684&131,548\\

FC layer &1,048,576&524,288&262,144&131,072\\

Proportion &99.91\%&99.87\%&99.79\%&99.64\%\\
\hline \hline
\end{tabular}
\end{table}

The CsiNet+ encoder is mainly composed of two convolutional layers, two BN layers, and one FC layer.
$N_{\rm conv} $, $N_{\rm BN}$, and  $N_{\rm FC}$ represent the parameter number of the convolutional layer, BN layer, and FC layer, respectively, which are calculated by,
\begin{equation}
N_{\rm conv}  = C_{\rm in}(K^2 + 1) C_{\rm out},
~
N_{\rm BN}  = 4 C_{\rm out},
~
N_{\rm FC}  = N_{\rm out}(N_{\rm in}+1) ,
\end{equation}
where $C_{\rm in}$ and $C_{\rm out}$ are the numbers of the input and output features of convolutional layer,  $K$ denotes the convolutional kernel size, and $N_{\rm out}$ and $N_{\rm in}$ represent the numbers of the input and output neurons of the FC layer, respectively.
As shown in Table \ref{NumMR}, the FC layer contains almost all model parameters that consume the most memory.
The number of the parameters for the FC layer at the fourfold compression encoder module is 1,048,576 and occupies 99.9\% of 1,049,500 overall parameters.
Therefore, it is critical to use the multiple-rate compression framework to decrease the parameter number of the FC layers.
We will reuse FC layers to decrease the encoder parameters.
There are two kinds of multiple-rate compression:
 \underline{s}eries \underline{m}ultiple-rate framework (SM-CsiNet+) and \underline{p}arallel \underline{m}ultiple-rate framework (PM-CsiNet+).
In the following, we will introduce SM-CsiNet+ and PM-CsiNet+ in detail.
\subsection{Series Multiple-Rate Compression Framework: SM-CsiNet+}
In general, highly compressed measurement vectors can be generated from the low ones, as in Fig. \ref{series}.
For instance, we can first compress the CSI matrix by fourfold and then continue to compress the compressed CSI matrix by twofold to obtain eightfold compression.
This method decreases the FC layer parameter number from $2048\times256$ to $512\times256$ for eightfold compression compared with compressing from the original CSI matrix.
Meanwhile, the first two convolutional layers, which are used to extract features, are also shared by different compression encoders, thereby further decreasing the number of encoder parameters.
Similarly, if we want to compress CSI by 16-fold, then we can keep on compressing the aforementioned compressed vector by twofold.

We train this multiple-rate compression framework by an end-to-end approach.
We concatenate the output of different decoders, generating a $32\times32\times8$ matrix as the output of the entire framework.
The label of this framework is obtained by repeating the original CSI matrix by fourfold.
We still use MSE as the loss function, which is calculated as follows,
\begin{equation}
\label{scale}
L_{\rm Total}(\Theta) = c_{\rm 4} L_{\rm 4}(\Theta_{\rm 4}) + c_{\rm 8} L_{\rm 8}(\Theta_{\rm 8}) + c_{\rm 16} L_{\rm 16}(\Theta_{\rm 16})  +   c_{\rm 32} L_{\rm 32}(\Theta_{\rm 32}),
\end{equation}
where $L_{\rm N}$, $c_{\rm N}$, and $\Theta_{\rm N}$ are the MSE loss, weight, and the learnable parameters of the $N$-fold compression network.
We can balance the magnitude of loss terms into similar scales by setting hyperparameter $c_{\rm N}$.
In the practical environments, the UE selects a suitable $CR$, and then the encoder compresses the CSI matrix to generate the corresponding measurement vectors.
Once the BS receives these measurement vectors, it decompresses them using the corresponding decoder network.

The parameter number of SM-CsiNet+ at the UE is 1,221,532, while that of the methods using different encoders to realize different $CR$s is 1,968,688.
Our proposed series framework decreases the parameter number by approximately 38.0\%, thereby greatly saving the storage space of the UE.
Meanwhile, if the more $CR$s need to be realized, then the parameter number reduction at the UE will be larger.

\subsection{Parallel Multiple-Rate Compression Framework: PM-CsiNet+}
\label{Parallel}
Although the series multiple-rate compression framework, SM-CsiNet+ in Fig. \ref{series}, greatly decreases the parameter number at the UE, it still occupies more storage space than the fourfold compression encoder.
Here, we develop a parallel multiple-rate compression framework that is with the same parameter number as fourfold compression encoder.

\begin{figure*}[t]
    \centering 
     \includegraphics[scale=0.56]{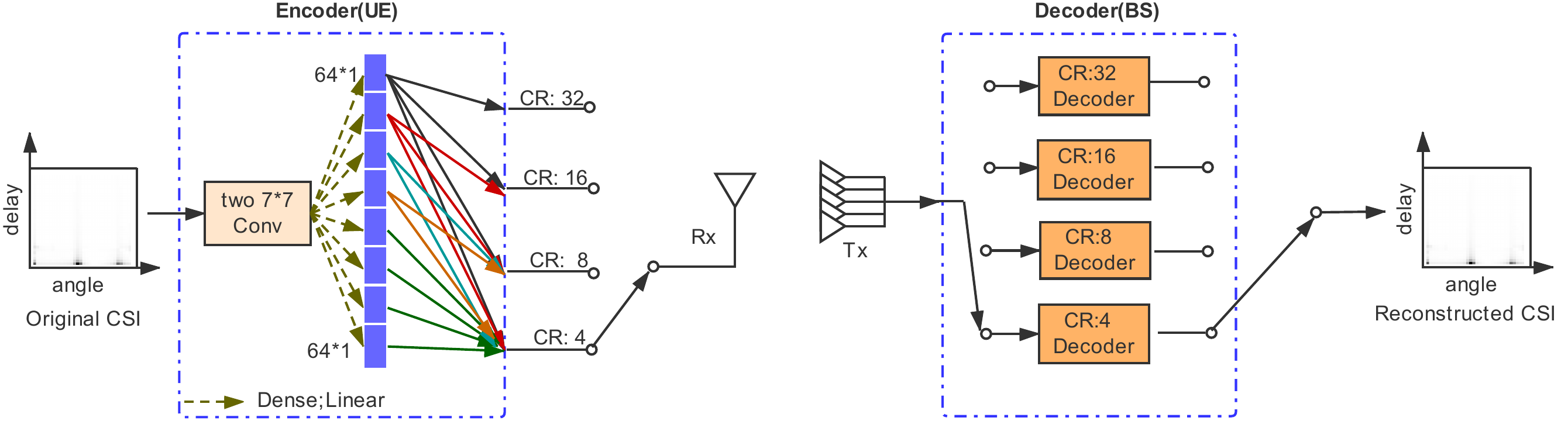}    
	\caption{\label{parallel}Parallel multiple-rate compression framework.} 
\end{figure*}

The key idea of the proposed SM-CsiNet+ is to generate measurement vectors with a large $CR$ from that with a small $CR$.
By contrast, parallel compression framework first compresses the CSI matrix with a large $CR$ many times and then generates measurement vectors with a small $CR$ via connecting those with large $CR$s in turn, as shown in Fig. \ref{parallel}.
For instance, the size of 16-fold measure vectors is $128\times1$, and that of 32-fold measurement vectors is $64\times1$, which is half of 16-fold measurement vectors.
Therefore, we can generate 16-fold measurement vectors via compressing CSI matrix by 32-fold twice and then connecting the measurement vectors together.
Similarly, an eightfold measurement vectors can be generated from two 16-fold measurement vectors or four 32-fold measurement vectors.

From another perspective, parallel framework can be also regarded as generating measurement vectors with a large $CR$ from that with a small $CR$. 
The encoder of this framework is the same as that of fourfold compression.
Specially, we first compress CSI matrix by fourfold and then select the part of the measurement vectors to generate vectors with larger $CR$s.
For example, when we need to compress CSI matrix by 32-fold, we just select the first 64 elements in the compressed vectors of fourfold compression.
Similarly, when 16-fold compression is needed, only the first 128 elements in the compressed vectors are selected.

Different from the rate-adaptive compressive sensing neural networks in \cite{lohit2018rate} using a three-stage training strategy, we still use an end-to-end approach to train parallel framework, similar to series framework.
Meanwhile, the loss function here is also the same as that of series framework.

The parameter number of this framework at the UE is 1,049,500 and the same as that of fourfold encoder.
This framework decreases parameter number by approximately 46.7\%, thereby greatly saving the storage space at the UE.

\section{Simulation Results and Discussions}
\label{results}
In this section, we first describe the details of our experiments.
Then, we evaluate the effects of the proposed two modifications on the reconstruction accuracy, compare with the existing state-of-the-art methods, and analyze the compression and reconstruction mechanism of CsiNet+.
Next, we evaluate the accuracy of the proposed quantization framework.
Finally, we analyze the two proposed multiple-rate compression frameworks.

\subsection{Experimental Setting}
\label{setting}
\subsubsection{Data generation}
We use the same dataset \footnote{\url{https://drive.google.com/drive/folders/1_lAMLk_5k1Z8zJQlTr5NRnSD6ACaNRtj?usp=sharing}} generated in \cite{8322184} to fairly compare CsiNet+ with CsiNet.
Here, we test the proposed networks and frameworks under COST 2100 MIMO channel model\cite{liu2012cost}.
The dataset includes two representative types of CSI matrices, namely, the indoor and outdoor rural scenarios, which are at the carrier frequency of 5.3 GHz and 300 MHz, respectively.
There are $N_c=1024$ subcarriers and $N_t = 32$ uniform linear array (ULA) antennas at the BS, respectively.
The complex CSI matrix in the angular-delay domain is first truncated to $32\times32$.
The other parameters are the same as \cite{liu2012cost}.

Here, we use no k-fold cross-validation because the dataset can be manually created without size limitation.
The  generated datasets are randomly divided into three parts, namely, training, validation, and testing sets,with 100,000, 30,000, and 20,000 samples, respectively.
During the experiment, the training set is used to update the model parameters.

\subsubsection{Hyperparameter setting}
\label{Hyperparameter}
Network models are initialized by a truncated normal initializer and optimized by the Adam optimizer\cite{kingma2014adam}.
CsiNet+ and two multiple-rate frameworks, SM-CsiNet+ and PM-CsiNet+, are all trained from the scratch.
Meanwhile, bit-level CsiNet+ is fine-tuned from those without quantization.
The batch sizes are all 200, while the epoch of the former three models and that of the latter model are 1000 and 200, respectively.
Moreover, the initial learning rate of the former models is 0.001 and that of the latter is 0.0001.
The learning rate will decay by half if the loss does not decrease in 20 epochs. 
Variables $c_{\rm 4}$, $c_{\rm 8}$, $c_{\rm 16}$, and $c_{\rm 32}$ are 30, 6, 2, and 1, respectively.
\subsubsection{Evaluation metric}
We utilize normalized MSE (NMSE) to measure CSI reconstruction accuracy, which is calculated as follows:
\begin{equation}
\rm NMSE =   \rm E \{ \|  \mathbf{H} -   {\mathbf{\hat H}}\|^2_2  /  \|  \mathbf{H}\|^2_2 \}.
\end{equation}

\subsection{Performance of the Proposed CsiNet+}
In this subsection, we first study the effects of two modifications on CSI reconstruction accuracy.
Then, we compare the performance of CsiNet+ with the state-of-the-art model CsiNet.
Finally, we visualize the parameters of the FC layer at the UE to explain the mechanism of the compression and reconstruction.
\subsubsection{Effect of modification 1}
\label{Effect of Modification 1}
\begin{table}[b]
\caption{NMSE ($\rm dB$) performance of Reconstructed CSI}
\label{modification}
\centering
\begin{tabular}{c|c|cccc}
\hline
\hline
                                               & CR  &CsiNet & CsiNet-M1 & CsiNet-M2  \\ \hline
\multirow{4}{*}{\rotatebox{90}{Indoor}}                        & 4  & -17.36 & -20.80     & -24.80       \\ 
                                               & 8  & -12.70 & -14.52     & -15.23        \\ 
                                               & 16 & -8.65                       & -11.77     & -12.21      \\ 
                                               & 32 & -6.24                       & -8.75      & -8.65        \\ \hline  \hline
                                               
\multirow{4}{*}{\rotatebox{90}{Outdoor}}                        & 4  & -8.75                       & -10.14     & -10.78    \\ 
                                               & 8  & -7.61                       & -8.11      & -8.55       \\  
                                               & 16 & -4.51                       & -4.99      & -4.44         \\  
                                               & 32 & -2.81                       & -1.87      & -2.78     \\ \hline \hline
\end{tabular}
\end{table}
As mentioned in Section \ref{M1}, we replace small convolutional kernels with $7\times7$ filters, thereby making full use of CSI block sparsity in the angular-delay domain.
Table \ref{modification} lists the performance of CsiNet-M1.
Evidently, CsiNet-M1 outperforms CsiNet in both indoor and outdoor scenarios.
However, the error reduction of outdoor scenario is much smaller than that of indoor scenario.
Although CSI matrices in both indoor and outdoor are sparse in the angular-delay domain, the outdoor CSI matrix exhibits much smaller `blank' area than the indoor CSI matrix.
The large kernel mainly works in the large `blank' area, thereby leading to the relatively smaller performance improvement in the outdoor scenario.

\subsubsection{Effect of Modification 2}
\label{Effect of Modification 2}
The motivation of modification 2 is from the refinement theory.
Table \ref{modification} shows the performance of CsiNet-M2.
From the table, CsiNet-M2 outperforms not only CsiNet but also CsiNet-M1, which demonstrates that the proposed modification efficiently improves the refinement performance.
Moreover, the performance improvement evidently decreases with the increase of $CR$ since the information loss of high $CR$ compression is unable to be offset even though the refinement has been strengthened.

\subsubsection{Comprehensive performance of CsiNet+}
Here, we compare our proposed CsiNet+ with the state-of-the-art CsiNet in reconstruction accuracy and parameter number.
\begin{figure}[!t]
    \centering 
     \includegraphics[scale=0.7]{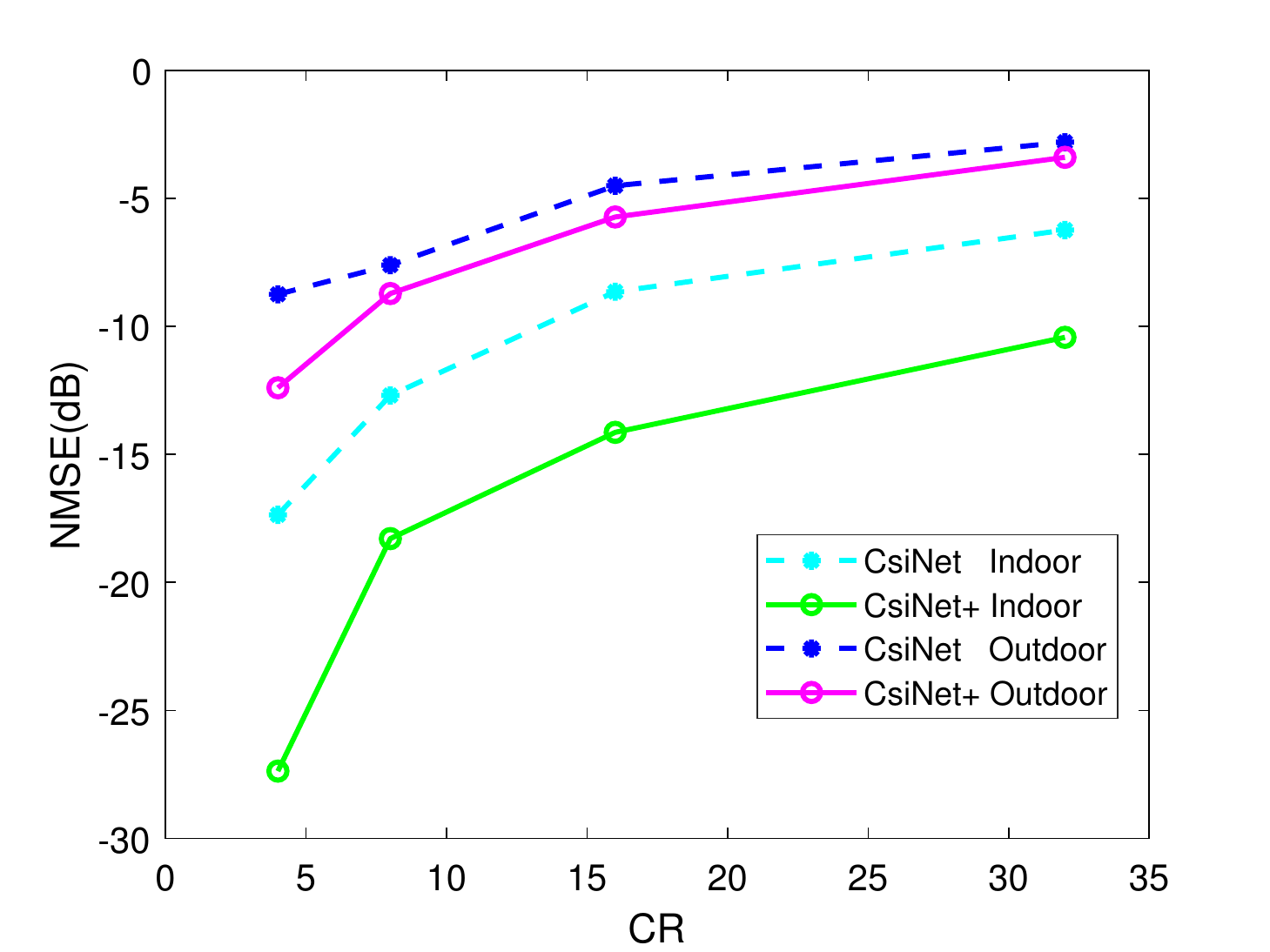}       
	  \caption{\label{CSINET+NMSE}$NMSE$ (dB) performance comparison between CsiNet+ and CsiNet. CsiNet+ shows noticeable accuracy advantages under all $CR$s.} 
\end{figure}

\begin{table}[t]
\caption{\label{CSINET+NUM}Parameter number comparison between CsiNet+ and CsiNet. }
 \centering 
\begin{tabular}{c|cccc}
\hline \hline
   \diagbox{Number}{$CR$}      & 4         & 8         & 16      & 32      \\  \hline
CsiNet   & 2,103,904 & 1,055,072 & 530,656 & 268,448 \\
CsiNet+  & 2,122,340 & 1,073,508 & 549,092 & 286,884 \\
Increase & 0.88\%    & 1.7\%     & 3.5\%   & 6.9\%  \\
\hline \hline
\end{tabular}
\end{table}
CsiNet+ shows noticeable accuracy advantages under all $CR$s, especially on small $CR$s, while the parameter number of the networks slightly increases, as shown in Fig. \ref{CSINET+NMSE} and Table \ref{CSINET+NUM}.

We use fourfold compression as an example and compare the time complexity of CsiNet and CsiNet+.
The processing time of CsiNet+ is 0.12 ms when tested on our 1080Ti GPU, while CsiNet requires 0.07 ms.
Although the parameter numbers are similar to CsiNet, the processing time greatly increases since the processing time is not only affected by parameter numbers but also dependent on the floating point operations (FLOPs)\cite{molchanov2016pruning}.
Due to the increase of convolutional kernel size in CsiNet+ from $3\times 3$ in CsiNet to $7\times 7$ and $5\times 5$, the FLOPs of CsiNet+ are significantly increased.
Although the processing time has greatly increased, it still meets the requirement of practical CSI reconstruction.

\subsubsection{Network robustness to channel noise}
\begin{figure*}[t]
    \centering 
    \subfigure [Indoor]{
     \includegraphics[scale=0.7]{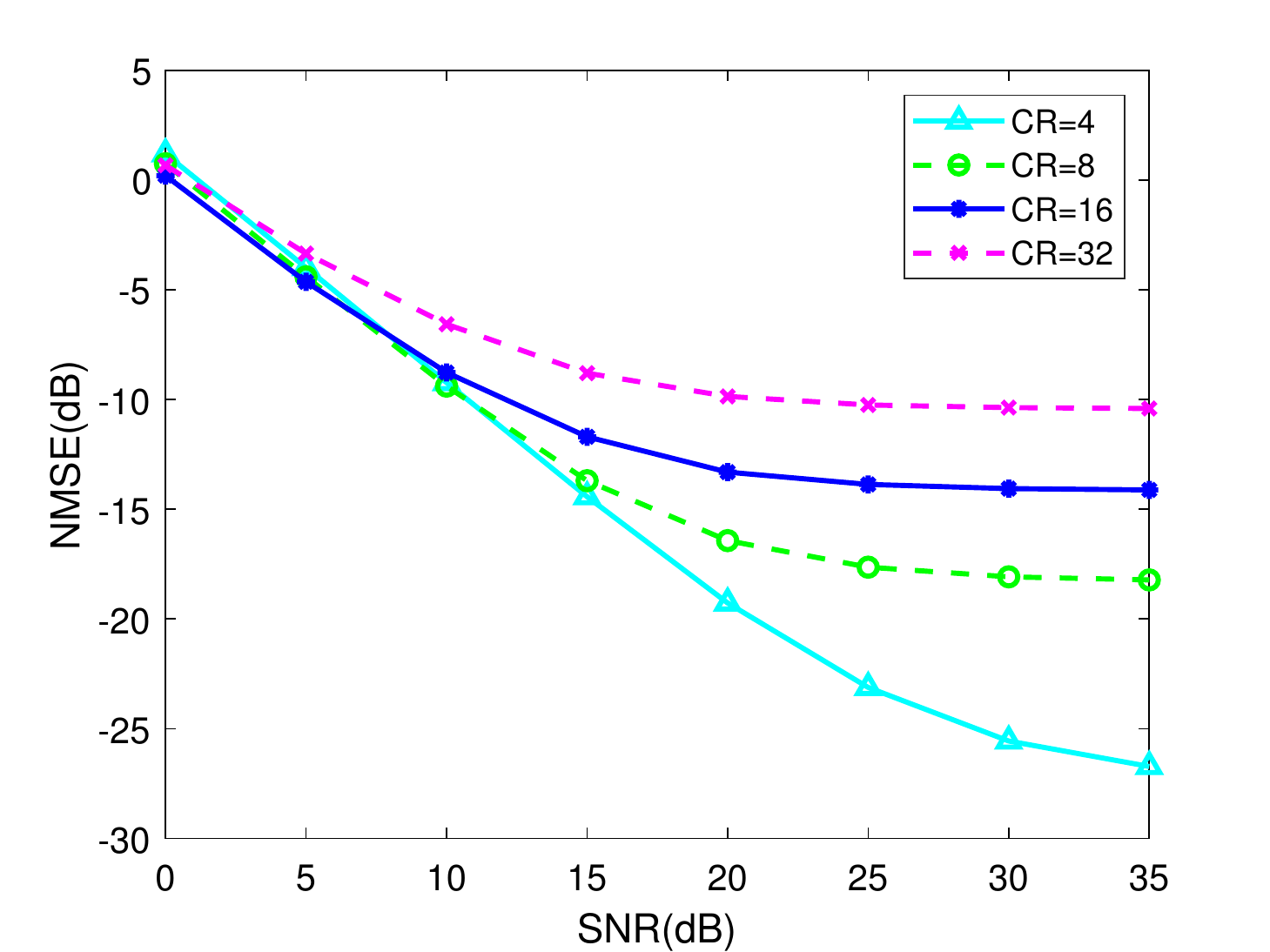}}   
     \subfigure [Outdoor]{
     \includegraphics[scale=0.7]{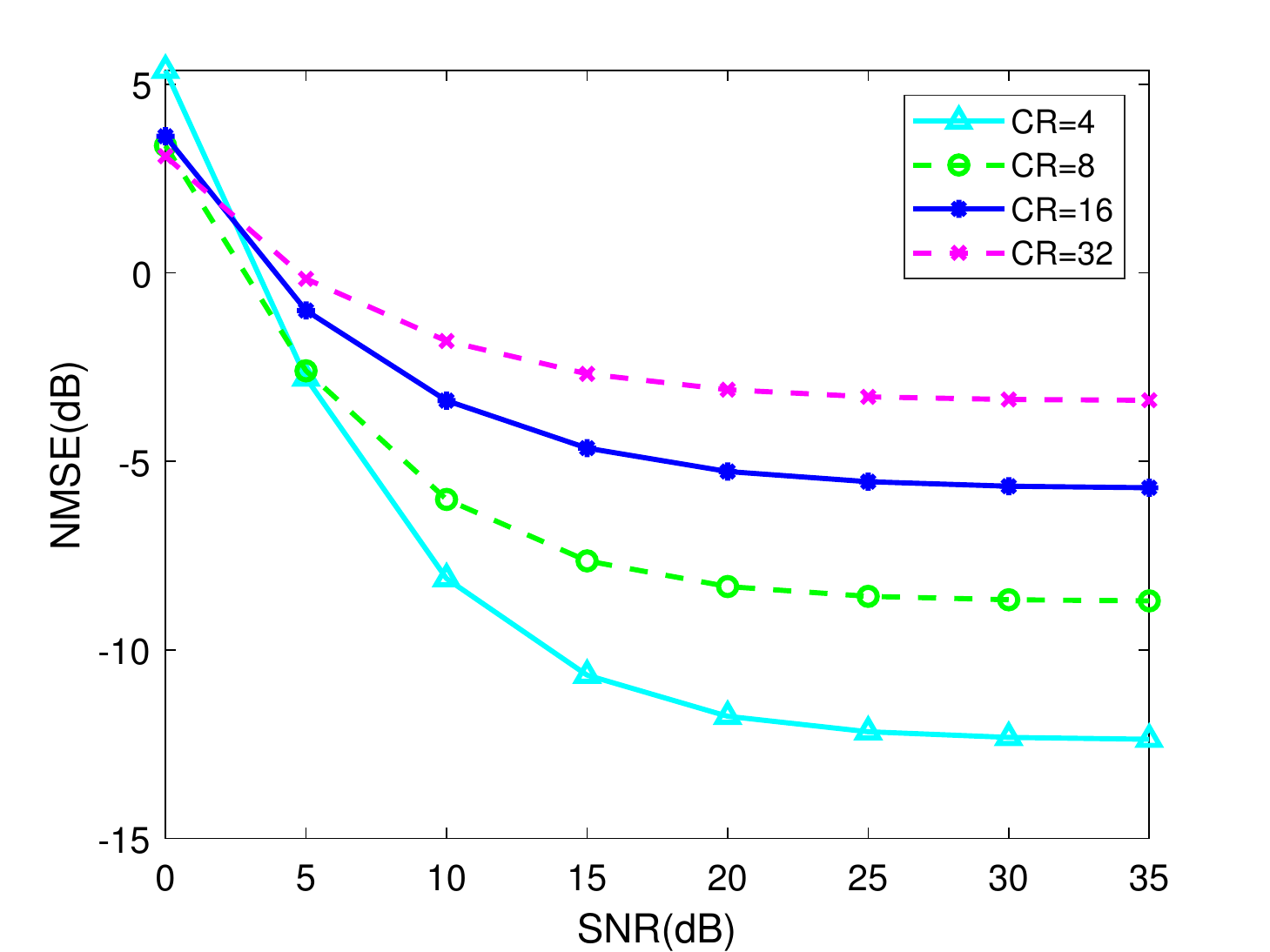}}   
        
	  \caption{\label{noiseUL}$NMSE$ (dB) vs SNR for the two scenarios under all $CR$s.} 
\end{figure*}
To test the robustness of the CsiNet+, we examine the CSI reconstruction performance in the presence of noise at the feedback channel.
Specially, we add an additive white Gaussian noise (AWGN) that corrupts the feedback information of measurement vectors.
We do not consider channel noise during network training because the DL-based methods perform almost the same when they are trained at an uplink SNR that is different from the actual operating uplink SNR \cite{jang2019deep}.
In Fig. \ref{noiseUL}, the NMSE performance of two scenarios is shown versus the uplink SNR.
High compression is more sensitive to the uplink SNR than low compression in both indoor and outdoor scenarios.
When the uplink SNR is below 5 dB, the $CR$ has little effect on the reconstruction accuracy because the DL-based method is able to learn statistical CSI automatedly and provide an initial estimate, as described in Section \ref{LearnEnvironments}.

\subsubsection{Neural network capacity}
\begin{figure}[t]
    \centering 
     \includegraphics[scale=0.7]{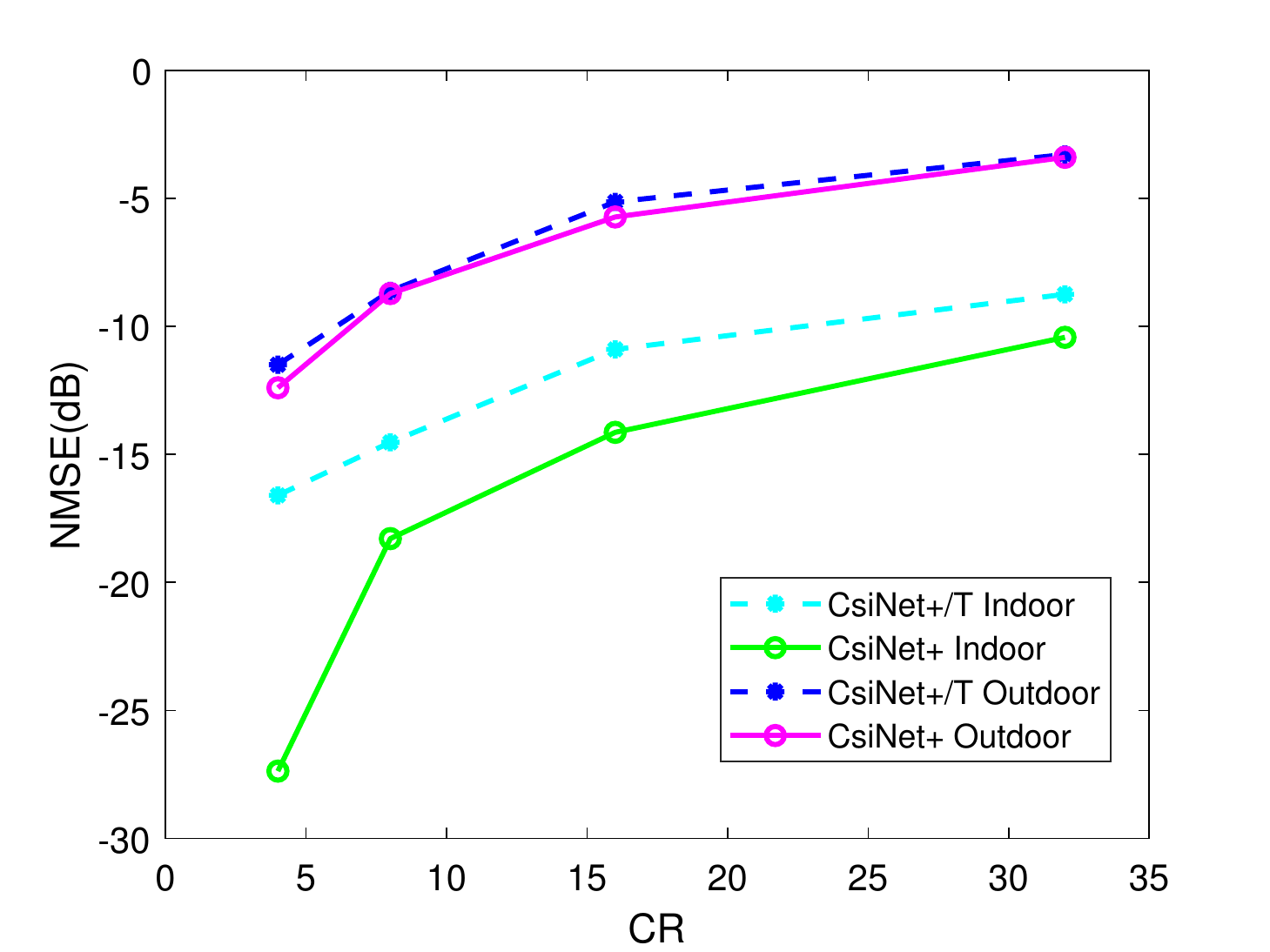}       
	  \caption{\label{1for2}$NMSE$ (dB) performance comparison between CsiNet+ and CsiNet+/T. CsiNet+/T denotes the method that uses a neural network to handle two scenarios simultaneously.} 
\end{figure}
In practical applications, UE will not always stay at a scenario and it may move from indoor to outdoor.
We investigate whether a neural network can handle two or more different scenarios simultaneously.
To test CsiNet+'s reconstruction performance for two scenarios, we use both of indoor and outdoor CSI to train CsiNet+.
From Fig. \ref{1for2}, the reconstruction accuracy of the indoor scenario CSI decreases a lot but there is little accuracy reduction of the outdoor scenario CSI reconstruction.
During training, the loss of the outdoor CSI is much larger than that of the indoor CSI and the MSE loss function primarily  smoothes the outdoor CSI reconstruction error, ignoring the indoor CSI\cite{guo2018relative}.

Although the reconstruction accuracy is not as high as before, this method has proved
the feasibility of DL-based CSI feedback operating at different scenarios.
The proposed CsiNet+ is of high expressivity by widening convolutional layers and deepening the network\cite{raghu2017expressive}, making it possible for a single model to reconstruct CSI of different scenarios.
Similarly, \cite{zhang2017beyond} also finds a single denoising CNN model can yield excellent results for three general image denoising tasks, i.e., super-resolution, blind Gaussian denoising, and JPEG deblocking, because of its high capacity.

\subsubsection{Compression and reconstruction mechanism}
\label{Commechanism}
We will map the FC layer parameters to the output of the former layer to observe the compression  process.
We use 32-fold compression as an example.
First, we reshape the FC layer parameter $\Theta_{\rm{fc}} \in  \mathbb{R}^{2048\times 64} $ to $2\times 32\times32\times 64$.
Then, we calculate the average of the absolute value of $\Theta_{\rm{fc}}$ over Axis 4, thereby obtaining a $2\times32\times32$ matrix $\Theta_{\rm{fc}}'$.
The normalized absolute FC layer parameters $\Theta_{fc}'$ are regarded as the values of heatmaps.
The larger the values are, the more attention the FC layer gives to the corresponding area.
\begin{figure*}[t]
    \centering 
    \subfigure [$CR=4$]{
     \includegraphics[scale=0.40]{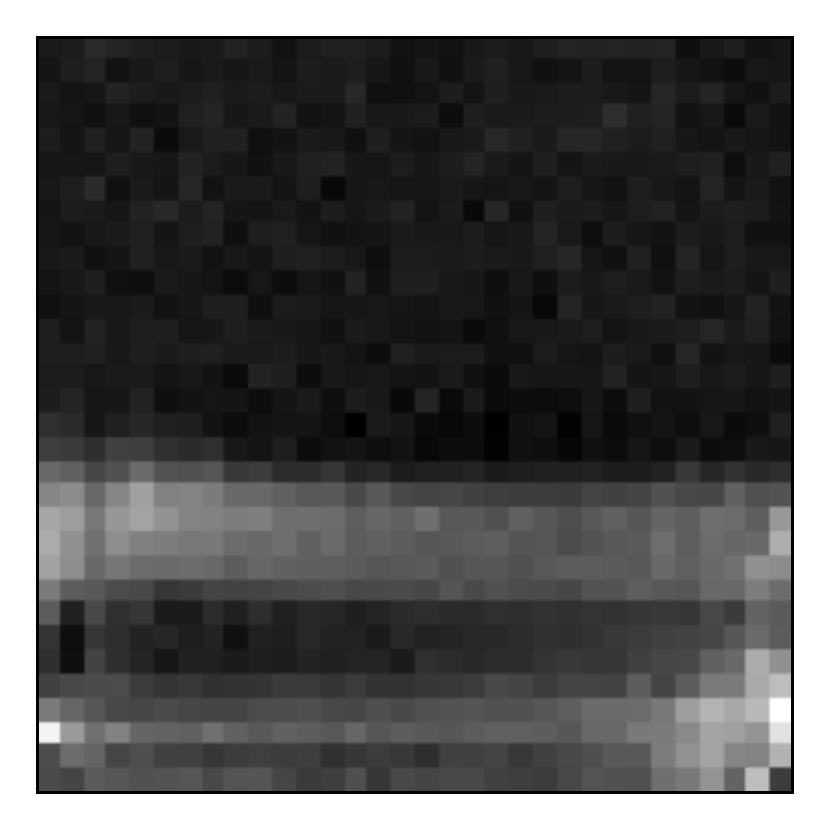}}   
    \subfigure [$CR=8$]{
     \includegraphics[scale=0.40]{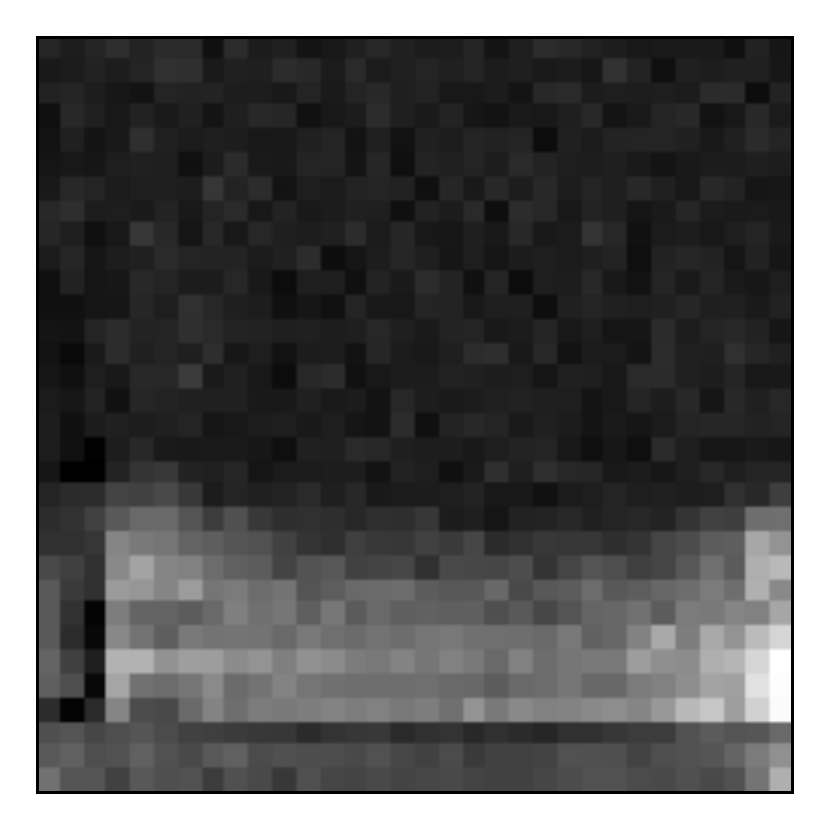} }   
\subfigure [$CR=16$]{
     \includegraphics[scale=0.40]{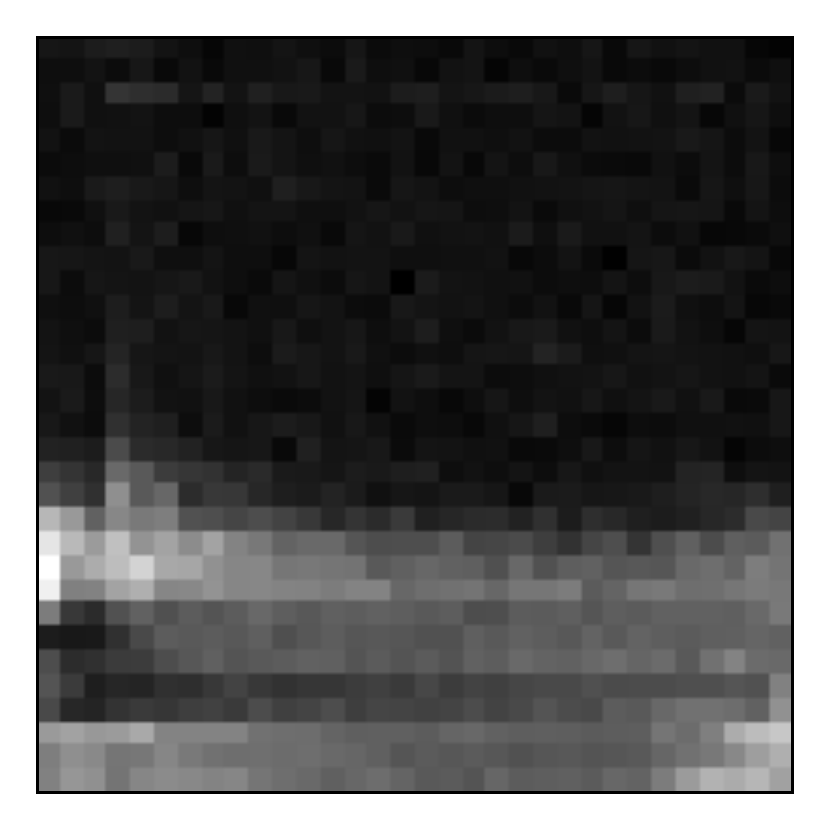}}   
    \subfigure [$CR=32$]{
     \includegraphics[scale=0.40]{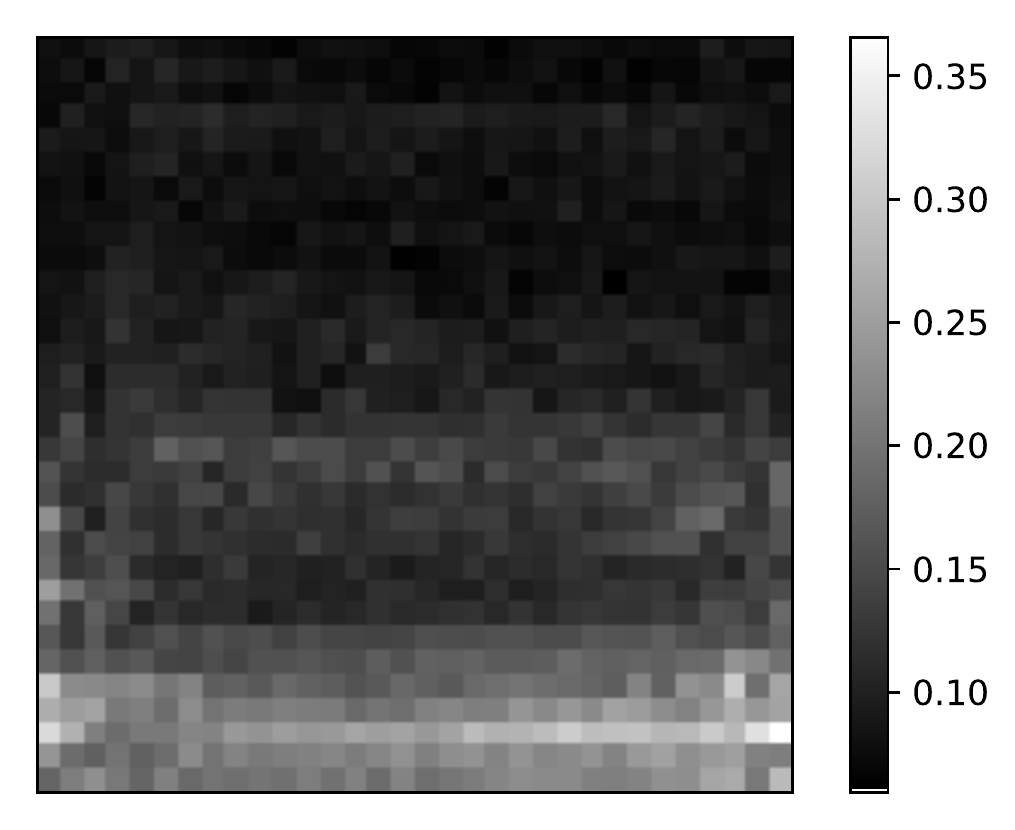} }   

\subfigure [$CR=4$]{
     \includegraphics[scale=0.40]{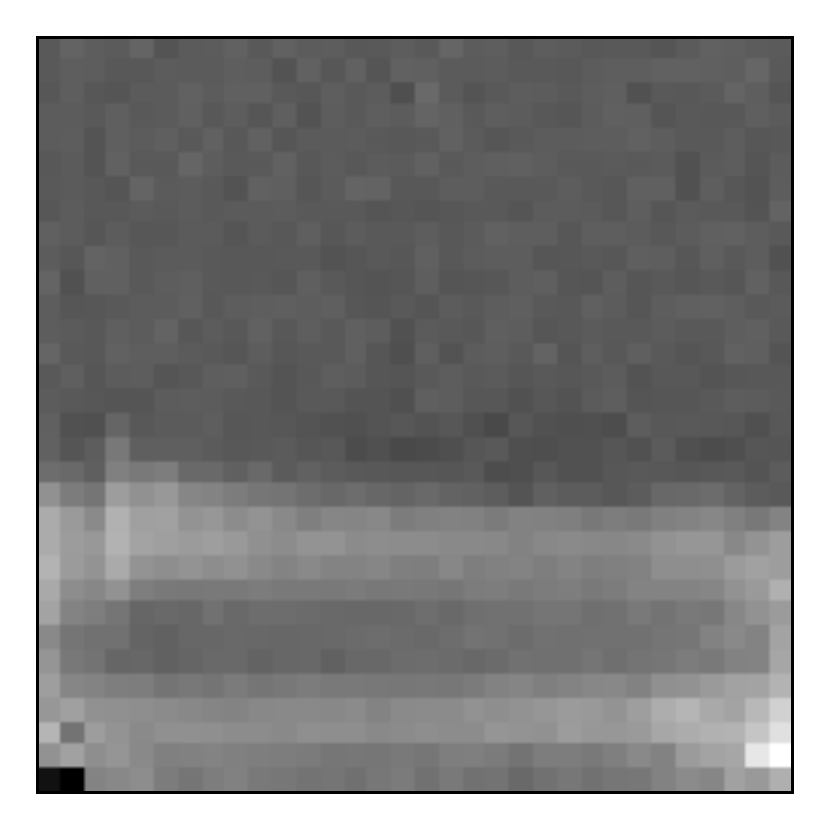}}   
    \subfigure [$CR=8$]{
     \includegraphics[scale=0.40]{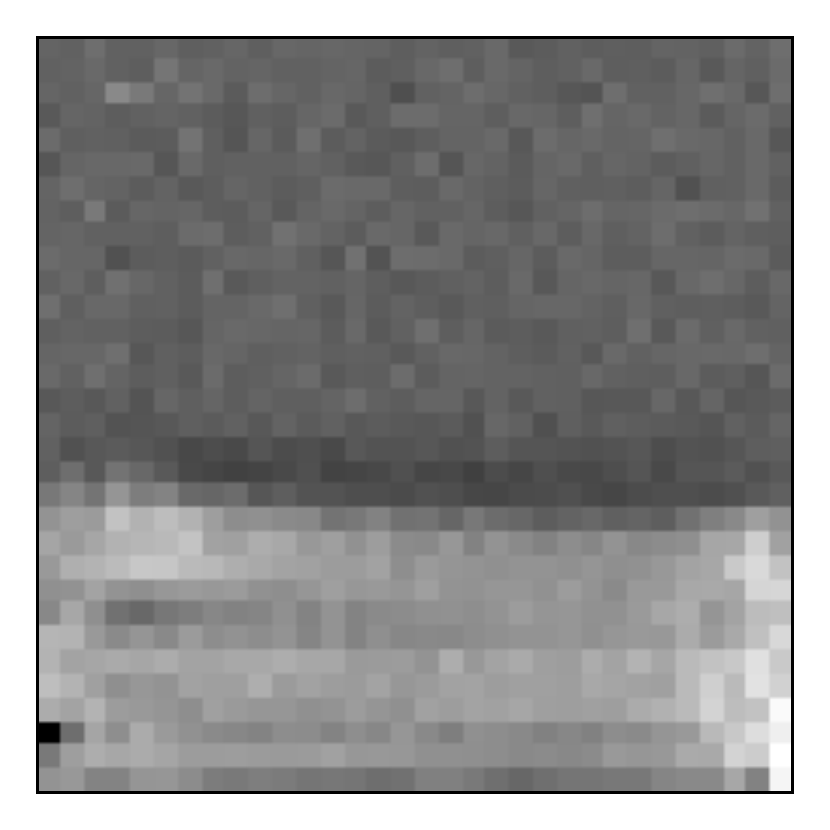} }   
    \subfigure [$CR=16$]{
     \includegraphics[scale=0.40]{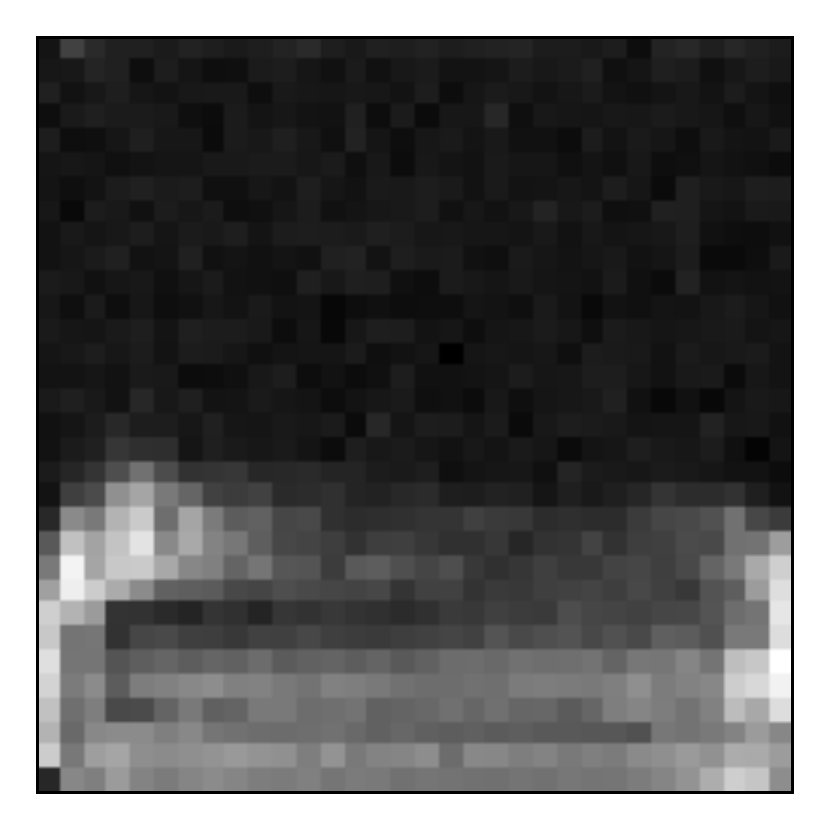}}   
    \subfigure [$CR=32$]{
     \includegraphics[scale=0.40]{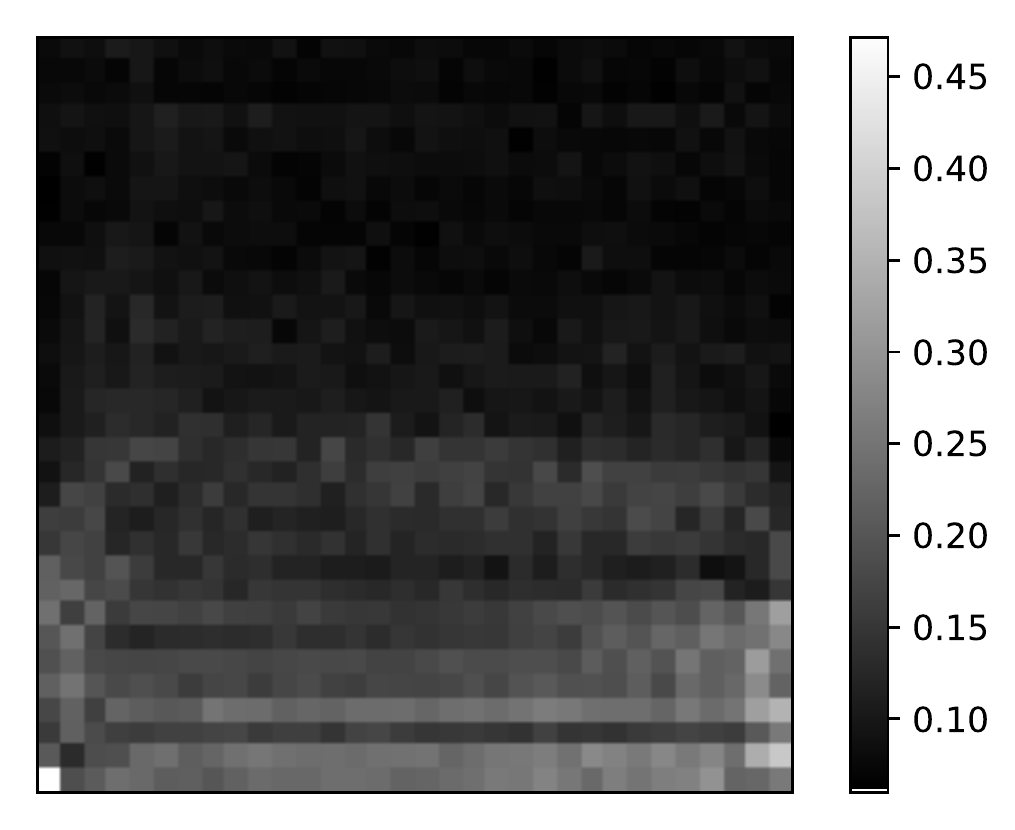} }   
	  \caption{\label{heatmap}FC layer parameter heatmaps of the indoor scenario for different $CR$s. The larger the values are, the more attention the FC layer gives to the corresponding area. Each FC layer exhibits two heatmaps, namely, the upper and bottom rows, because the output of the convolutional layers at the encoder is $32\times32\times2$.} 
\end{figure*}
From Fig. \ref{heatmap} , the areas of great interest to the FC layers are the bottoms of the feature maps.
Meanwhile, the remaining area is full of near-zero values and contains little information.
Therefore, the DL-based CS feedback methods utilize FC layers to determine the non-zero areas and then generate measurement vectors by mainly exploiting the information of these areas.
With the increase of $CR$s, the areas of interest become smaller, and the upper areas are given little attention to, as shown in Fig. \ref{heatmap}.
Specifically, high compression is at the expense of the loss of incidental information.
Evidently, the reconstruction accuracy drops with the increase of $CR$. 

\begin{figure}[t]
    \centering 
    \subfigure []{
     \includegraphics[scale=0.40]{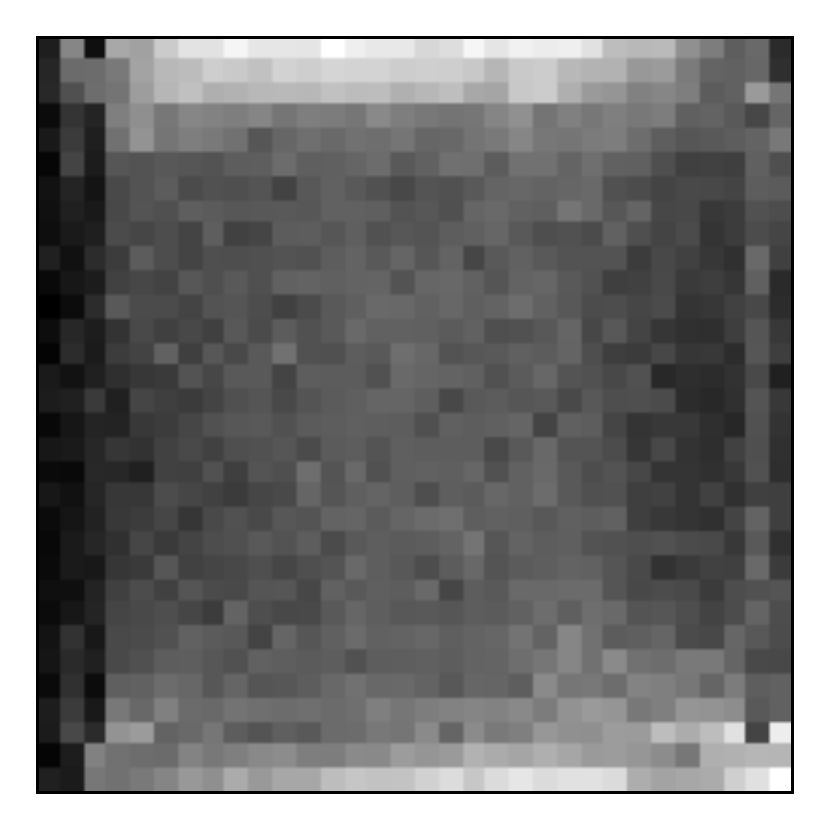}}   
    \subfigure []{
     \includegraphics[scale=0.40]{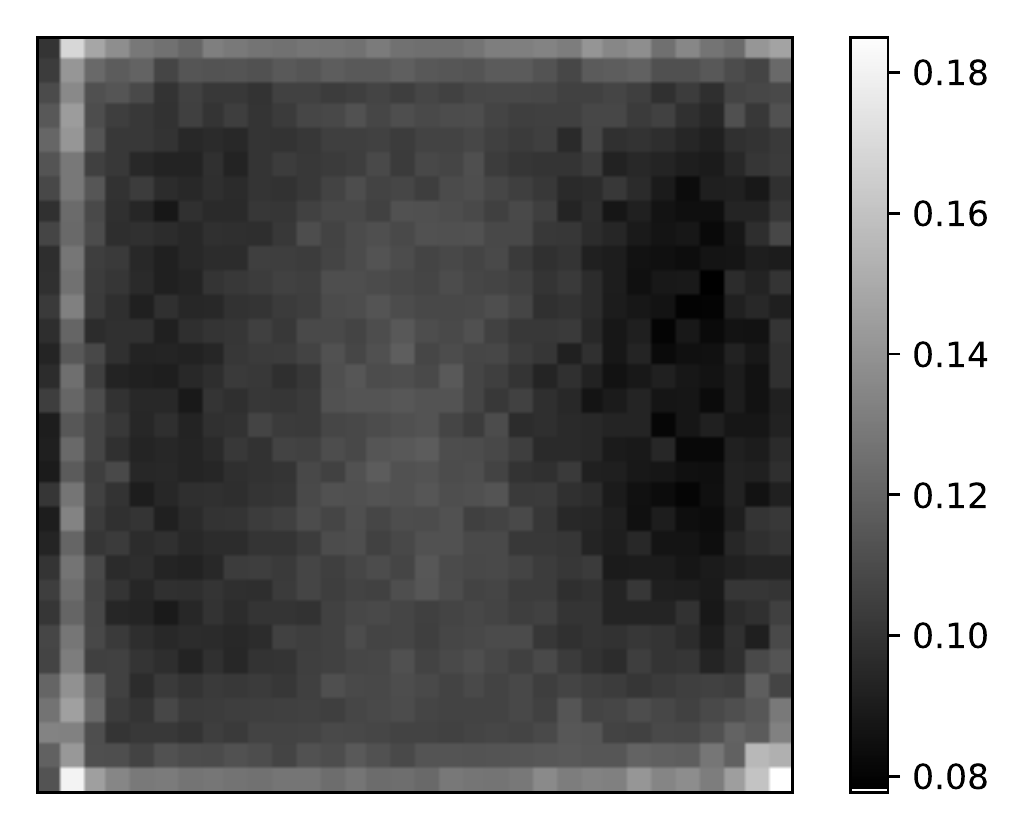} }     
	  \caption{\label{heatmapOut}FC layer parameter heatmaps of outdoor scenario for fourfold compression.} 
\end{figure}
In both CsiNet and CsiNet+, the reconstruction accuracy of the outdoor scenario is much lower than that of the indoor scenario. 
We can compare the heatmaps of indoor and outdoor scenarios in Fig. \ref{heatmap} and \ref{heatmapOut}. 
From Fig. \ref{heatmapOut}, CsiNet+ is unable to efficiently extract the information of the outdoor CSI since the outdoor CSI exhibits much little sparsity than the indoor CSI.

\begin{figure}[t]
    \centering 
\subfigure []{\label{environments1}
     \includegraphics[scale=0.40]{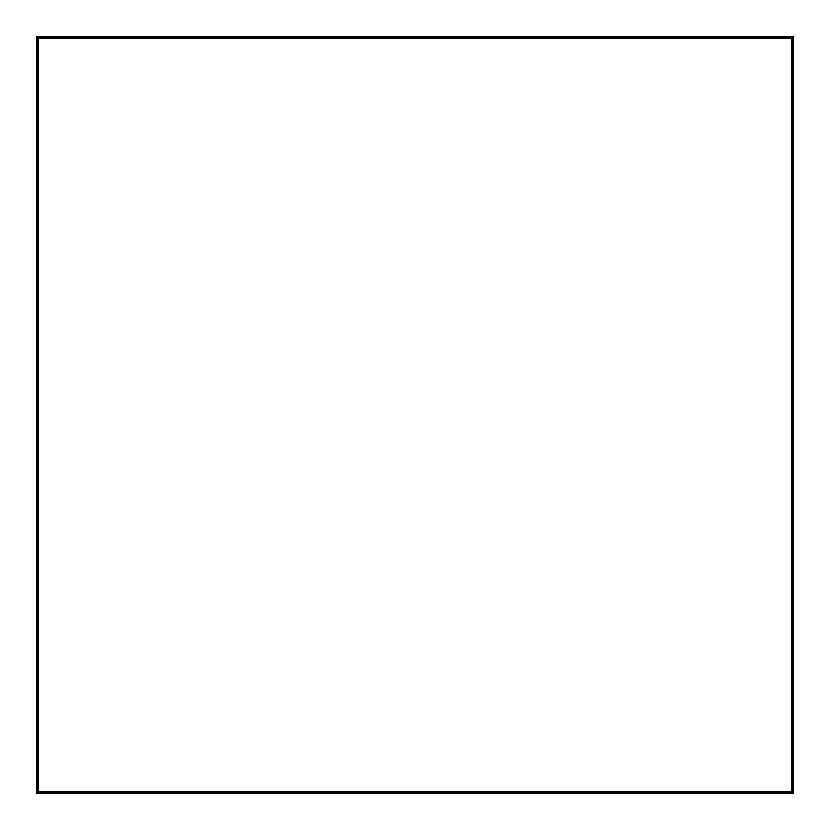}    }
\subfigure []{\label{environments2}
     \includegraphics[scale=0.40]{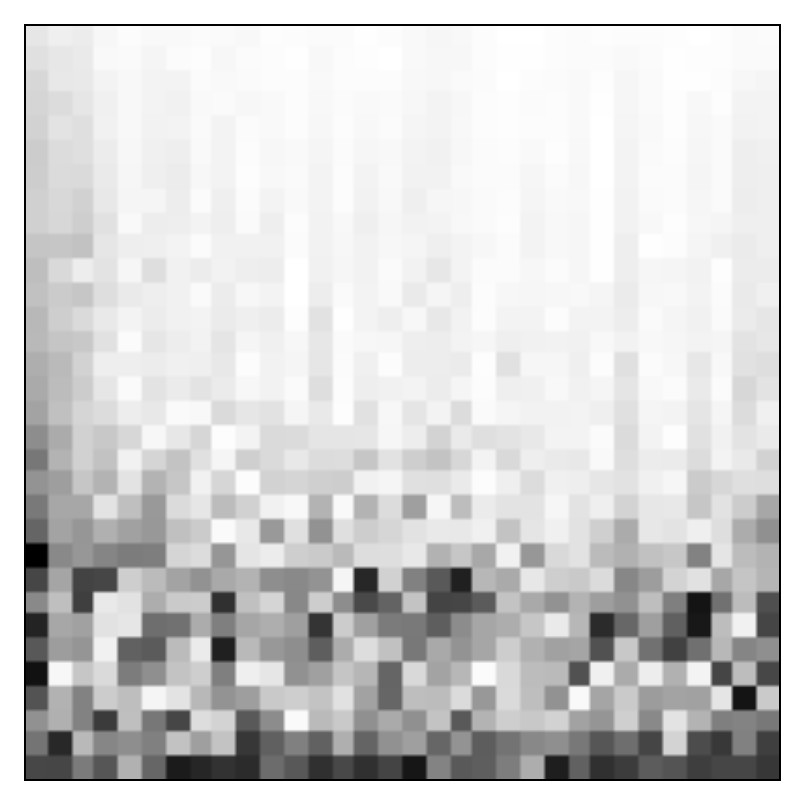}    }
\subfigure []{\label{environments3}
     \includegraphics[scale=0.40]{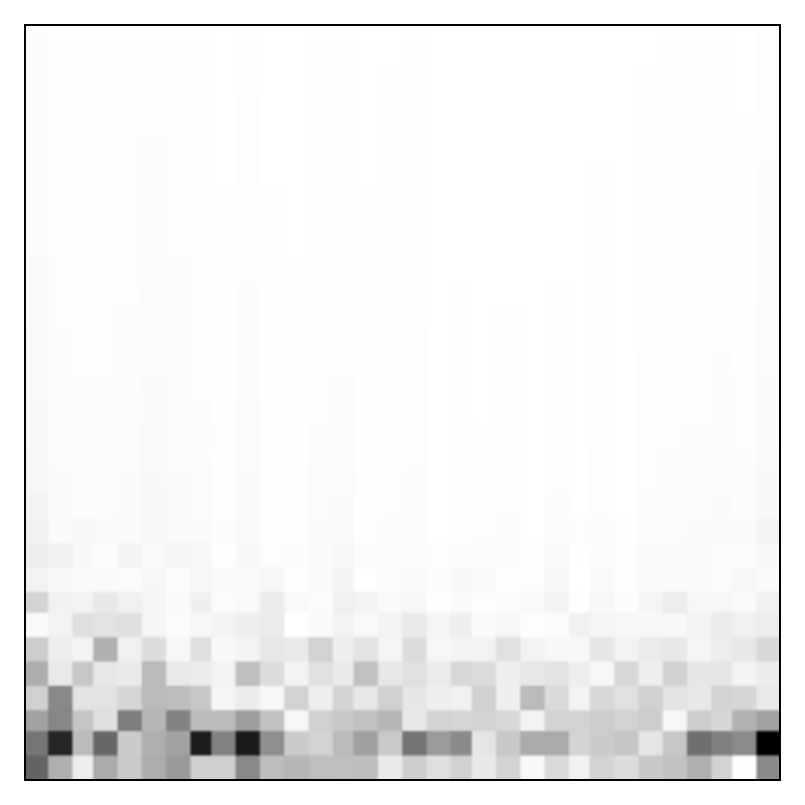}    }
	\caption{\label{environments}CsiNet+'s ability to learn statistical CSI. a: all-zero input; b: output of CsiNet+; c: mean of training dataset. Although the output and mean CSI are not visually exactly the same, some remarkable similarities can be observed.} 
\end{figure}

\label{LearnEnvironments}
The CS-based algorithms only use signal sparsity as prior information and neglect other signal characteristics.
DL-based methods can extract all useful features from data automatedly. 
The proposed CsiNet+ learns not only the sparse structure of CSI but also the statistical CSI.
We feed a $32\times32\times2$ all-zero matrix, as in Fig. \ref{environments1}, into the trained CsiNet+.
Fig. \ref{environments2} and \ref{environments3} are the corresponding output of CsiNet+ and the mean of CSI matrices at the indoor scenario, respectively.
Although the output and mean CSI are not visually exactly the same, some remarkable similarities can be observed.
The upper parts contain large `blank' areas while the bottoms contain the most information. 
Therefore, the DL-based CSI feedback method CsiNet+ can automatically learn statistical CSI.

The DL-based methods make use of not only limited feedback but also the statistical CSI learned automatedly.
Therefore, even when the feedback information is little or noisy, the DL-based methods can reconstruct high-quality CSI compared with traditional CS-based methods.

\subsection{Quantization Evaluation}
\begin{table*}[ht]
\centering
\caption{NMSE ($dB$) performance of the proposed quantization methods for different scenarios. Key for method abbreviations: UQ: uniform quantization; NUQ: non-uniform quantization; NUQ+O: non-uniform quantization with the offset network.}
\resizebox{\textwidth}{!}{
\begin{tabular}{c|cccc|cccc|cccc|cccc}  \hline \hline
 &\multicolumn{12}{c}{Indoor}                                                                                                                                                                                                                                                             \\  \hline 
\multicolumn{1}{c|}{$CR$  }                                                           & \multicolumn{4}{c|}{4}                               & \multicolumn{4}{c|}{8}                               & \multicolumn{4}{c|}{16}                              & \multicolumn{3}{c}{32}                             \\
\hline
\multicolumn{1}{c|}{$B$}            &3                                                  & 4               & 5               & \multicolumn{1}{c|}{6}           &3      & 4               & 5               & \multicolumn{1}{c|}{6}             &3   & 4               & 5               & \multicolumn{1}{c|}{6}              &3     & 4              & 5               & 6 \\
\hline
\multicolumn{1}{c|}{CsiNet+}&
\multicolumn{4}{c|}{-27.37}&\multicolumn{4}{c|}{-18.29} &\multicolumn{4}{c|}{-14.14} &\multicolumn{4}{c}{-10.43} \\ \hline
\multicolumn{1}{c|}{UQ}                                           &-12.97               & -16.78          & -20.35          & -23.53      &-11.33     & -14.39         & -16.22          & -17.79   &-9.64       & -12.11          & -13.43          & -13.93     &-7.12     & -8.93          & -9.96           & -10.29          \\
\multicolumn{1}{c|}{NUQ}                                    &-14.82                  & -18.55          & -22.00          & -24.95    &-12.45      & -15.39          & -17.20          & -18.02      &-10.35    & -12.56          & \textbf{-13.64} & \textbf{-14.04}& -7.82 & -9.36          & -10.09          & \textbf{-10.35} \\
\multicolumn{1}{c|}{NUQ+O} & \textbf{-15.27}&\textbf{-19.06} & \textbf{-22.13} & \textbf{-24.97} &\textbf{-12.81} & \textbf{-15.65} & \textbf{-17.23} & \textbf{-18.03}  &\textbf{-10.48} & \textbf{-12.58} & \textbf{-13.64} & -14.02   &\textbf{-7.91}       & \textbf{-9.37} & \textbf{-10.10} & \textbf{-10.35} \\
 \hline \hline
 
 & \multicolumn{12}{c}{Outdoor}                                                                                                                                                                                                                                                            \\  \hline 
\multicolumn{1}{c|}{$CR$  }                                                           & \multicolumn{4}{c|}{4}                               & \multicolumn{4}{c|}{8}                               & \multicolumn{4}{c|}{16}                              & \multicolumn{4}{c}{32}                             \\
\hline
\multicolumn{1}{c|}{$B$}          &3                                                      & 4               & 5               & \multicolumn{1}{c|}{6}         &3        & 4               & 5               & \multicolumn{1}{c|}{6}             &3  & 4               & 5               & \multicolumn{1}{c|}{6}          &3         & 4              & 5               & 6 \\
\hline 
\multicolumn{1}{c|}{CsiNet+}&
\multicolumn{4}{c|}{-12.40}&\multicolumn{4}{c|}{-8.72} &\multicolumn{4}{c|}{-5.73} &\multicolumn{4}{c}{-3.40} \\ \hline
UQ                              &-7.97                            & -10.37          & -11.66          & -12.18          &-6.51   & -7.87           & -8.43           & -8.65       &-4.05    & -4.93           & -5.41           & -5.61   &\textbf{-2.54  }      & \textbf{-2.99} & \textbf{-3.35}  & -3.34           \\

NUQ &-9.45 & -11.17          & -11.96          & -12.28       &-6.53   & -7.90           & -8.44           & -8.64     &-4.65       & -5.33           & -5.59           & -5.68     &-2.45      & -2.92          & -3.27           & -3.37           \\
NUQ+O&   \textbf{-9.96} &  \textbf{-11.54} & \textbf{-12.42} & \textbf{-12.83} & \textbf{-6.67}&  \textbf{-7.94}  & \textbf{-8.51}  & \textbf{-8.81}  & \textbf{-4.94}& \textbf{-5.49}  & \textbf{-5.70}  & \textbf{-5.75}  & -2.53  & -2.93          & -3.18           & \textbf{-3.79}  \\ 
\hline \hline
\end{tabular}}
\label{quanNMSE}
\end{table*}
Table \ref{quanNMSE} shows the NMSE performance of our proposed bit-level CsiNet+.
From the table, non-uniform quantization outperforms uniform quantization by a margain as we imagine because the measurement vectors are weak signals.
Non-uniform quantization with the offset network achieves the best quantization performance.
From the perspective of refinement theory, the stacked FC layers in the offset network refine the output of the dequantizer, thereby minimizing quantization distortion.

As we can imagine, the reconstruction is becoming more and more accurate with the increase of quantization bits.
Moreover, CsiNet+Q6 even exhibits a similar performance as the original CsiNet+ without quantization.
Second, the indoor scenario is more sensitive to the change of quantization bits compared with the outdoor scenario.
Similarly, low compression is much more sensitive than high compression.
This phenomenon may be attributed to that high-accuracy reconstruction is based on the full use of measurement vectors.
Therefore, little distortion of measurement vectors will lead to a large decrease in reconstruction accuracy.
Besides,  CsiNet+Q6 at the outdoor scenario even outperforms that with quantization because the encoder cannot efficiently extract the information of outdoor CSI and quantization drops some redundant information.  

In practical scenario, $CR$ and quantization bits $B$ together determine the overhead of CSI feedback.
For example, if the feedback bitstream contains 1536 bits, we can have two choices, $CR$=4,  $B$=3, or $CR$=8, $B$=6.
The NMSE of the former at the indoor scenario is --15.27 dB, while that of the latter is --18.03 dB, as shown in Table \ref{CRvsB}.
\begin{table*}[t]
\centering
\caption{NMSE($dB$) performance with fixed bitstream length}
\begin{tabular}{c|cc|cc|cc|cc|cc|cc} \hline \hline
Scenario   & \multicolumn{6}{c|}{Indoor}                                                   & \multicolumn{6}{c}{Outdoor}                                                  \\   \hline
Total bits & \multicolumn{2}{c|}{1536} & \multicolumn{2}{c|}{738} & \multicolumn{2}{c|}{384} & \multicolumn{2}{c|}{1536} & \multicolumn{2}{c|}{738} & \multicolumn{2}{c}{384} \\ \hline
$CR$/$B$   & 4/3         & 8/6        & 8/3        & 16/6       & 16/3       & 32/6       & 4/3         & 8/6        & 8/3        & 16/6       & 16/3       & 32/6       \\
NMSE($dB$) & -15.27      & -18.03     & -12.81     & -14.02     & -10.48     & -10.35     & -9.96       & -8.81      & -6.67      & -5.75      & -4.94      & -3.79     \\ \hline \hline
\end{tabular}
\label{CRvsB}
\end{table*}
From Table \ref{CRvsB}, at the indoor scenario, with a fixed bitstream length, compared with decreasing quantization bits, the increase of $CR$ has less bad effect on reconstruction accuracy.
In other words,  the CSI reconstruction accuracy is more senstive to quantization distortion.
At the outdoor scenario, to the opposite, the CSI reconstruction accuracy is more senstive to compression errors.
This gives a guideline to practical deployment that,  even if the length of feedback bitstream is fixed, suitable $CR$ and quantization bits must be selected to achieve optimal performance at the practical scenario, which also shows the necessity of the proposed quantization strategy and multiple-rate frameworks.

\subsection{Performance of Multiple-Rate Compression Framework}
\begin{table}[t]
\caption{NMSE ($\rm dB$) performance of the proposed SM-CsiNet+ and PM-CsiNet+}
\label{multiple}
\centering
\begin{tabular}{c|c|cccc}
\hline
\hline
                                               & $CR$  & CsiNet & CsiNet+ & SM-CsiNet+ & PM-CsiNet+ \\ \hline
\multirow{4}{*}{\rotatebox{90}{Indoor}}                           & 4  & -17.36 & -27.37  & \textbf{-27.90}      & -27.60      \\  
                                               & 8  & -12.70 & -18.29  & \textbf{-18.49}      & -17.70      \\   
                                               & 16 & -8.65  & \textbf{-14.14}  & -13.45      & -12.25      \\   
                                               & 32 & -6.24  & \textbf{-10.43}  & -9.89       & -8.24       \\ \hline \hline
                                                
{\multirow{4}{*}{{\rotatebox{90}{Outdoor}}   }} & 4  & -8.75  & \textbf{-12.40}  & -11.91      & -12.02      \\  
                         & 8  & -7.61  & \textbf{-8.72}   & -8.25       & -8.10       \\ \cline{2-6} 
                        & 16 & -4.51  & \textbf{-5.73}   & -5.31       & -5.07       \\ \cline{2-6} 
                       & 32 & -2.81  & \textbf{-3.40}   & -3.22       & -3.00       \\ \hline  \hline
\end{tabular}
\end{table}

Table \ref{multiple} shows the NMSE (dB) performance of the multiple-rate compression frameworks.
From the table, multiple-rate compression frameworks are not lossless compared with the direct compression network. 
The series framework, SM-CsiNet+, has similar reconstruction accuracy to the direct one, as shown in Table \ref{multiple}. 
When the $CR$ is 4 or 8 for the indoor scenario, the series compression network performs better than the direct one. 
In other cases, its accuracy is slightly worse than the direct one by approximately 0.5 dB. 
The parallel framework,  PM-CsiNet+, is approximately 2 dB worse than the direct one.

From Table \ref{multiple}, SM-CsiNet+ outperforms PM-CsiNet+ by a margin, because the parameter number of PM-CsiNet+ is approximately 85.9\% of that of SM-CsiNet+ and the stacked FC layers increase the depth of SM-CsiNet+.
In general, the DL-based methods exhibit enhanced fitting ability with the increase of parameter number and layer depth in neural networks\cite{lecun2015deep}.

Further more, SM-CsiNet+ even performs better than CsiNet+ at the indoor scenario when the $CR$ is four or eight, which can be explained by the regularization theory.
We use fourfold compression as an example.
In the testing period, SM-CsiNet+ for fourfold compression is the same as CsiNet+ in not only the network architecture but also the parameter number.
Therefore, the performance improvement of SM-CsiNet+ results from the training period rather than network architecture or parameter number.
If only focusing on fourfold compression during training, we can regard the subsequent compression networks as an additional regularization term\cite{8233175}. 
If the fourfold compression measurements are inefficient, other high-compression measurements generated from the fourfold compression measurements cannot contain enough useful information, thereby leading to the poor reconstruction accuracy of other high compression. 
Specifically, the subsequent compression networks force the former compression networks to extract useful information as much as possible, which is the reason behind the excellent performance of multiple-rate frameworks at low $CR$s.

As mentioned in (\ref{scale}) and Section \ref{Hyperparameter}, $c_{\rm 4}$, $c_{\rm 8}$, $c_{\rm 16}$, and $c_{\rm 32}$ are used to balance the magnitude of loss terms into similar scales and set as 30, 6, 2, and 1, respectively.
During the experiments, the performance changes of different compression sub-networks differ if we change these parameters.
For example, we change $c_{\rm 16}$ into 20 and 200, respectively.
Table \ref{gamma} shows that the reconstruction accuracy of 16-fold compression is improved, with the increase of $c_{\rm 16}$, because the increase of $c_{\rm 16}$ forces the training process to focus more on the 16-fold compression sub-network at the expense of other sub-networks' performance.
\begin{table}[t]
\caption{NMSE ($\rm dB$) performance of PM-CsiNet+ with different $c_{\rm 16}$ at outdoor scenario.}
\label{gamma}
\centering
\begin{tabular}{c|cccc}
\hline \hline
\diagbox{$c_{\rm 16}$}{$CR$}  & 4     & 8    & 16   & 32   \\ \hline
2        & -12.02 & -8.10 & -5.07 & -3.00 \\ 
20       & -11.78 & -7.90 & -5.39 & -2.98 \\ 
200      & -11.84 & -7.51 & -5.56 & -2.73 \\ \hline \hline
\end{tabular}
\end{table}

This makes the proposed multiple-rate frameworks more suitable for practical applications.
We can increase the weight of the preferred sub-network in (\ref{scale}) because the UE commonly possesses the preferred $CR$, which is decided by its communication environment. 

\section{Conclusion}
\label{Conclusion}
In this paper, we have proposed a multiple-rate bit-level compressive sensing DL-based framework for CSI feedback problem described in Section \ref{introduction}. 
We focus on the CSI reconstruction accuracy and feasibility at the UE.

We first introduce a network architecture, CsiNet+, modified from CsiNet.
In contrast with other DL-based methods, which regard neural networks as a black box and only care about the reconstruction accuracy, we explain the motivation of the modifications and the compression mechanism via parameter visualization of the FC layer and determine CsiNet+'s ability to learn statistical CSI automatedly.

We propose a novel framework and training strategy for quantization problem.
Other quantization strategies need to train different networks for distinct quantization bits.
By contrast, we do not need any parameter update at the UE, and just fine-tune the parameters in the decoder at the BS for different quantization bits.
Besides, we introduce an offset network to minimize quantization distortion.

Lastly, we propose two frameworks SM-CsiNet+ and PM-CsiNet+ to overcome the problem that DL-based methods need to store different parameter sets for distinct $CR$s, thereby leading to great storage space waste.
The proposed two frameworks reduce parameter number by 38.0\% and 46.7\% compared with the existing DL-based method  that train and store a different neural network for a different $CR$.
SM-CsiNet+ outperforms PM-CsiNet+ at the expense of increasing the parameter number.
SM-CsiNet+ even performs better than CsiNet+ at the low $CR$.

\ifCLASSOPTIONcaptionsoff
  \newpage
\fi

\bibliographystyle{IEEEtran}
\bibliography{IEEEabrv,reference}
\end{document}